\documentclass[epj]{svjour}
\usepackage{graphics}
\usepackage{hyperref}
\usepackage{titlesec}
\usepackage{graphicx,epsfig,epstopdf}
\usepackage{bm}
\usepackage{epstopdf}
\usepackage{amssymb,amsmath,amsfonts,latexsym}
\usepackage{dcolumn}
\usepackage{bm}
\usepackage{hyperref}
\usepackage[italic]{hepnames}
\usepackage[utf8]{inputenc}

\usepackage{breqn}
\usepackage{breqn}
\usepackage{amssymb}
\usepackage{multirow}
\usepackage{breqn}       % These four lines are
\makeatletter         % used to line-break equations
\let\cat@comma@active\@empty  % automatically
\makeatother          % use \begin{dmath} ... \end{dmath} for eqns

\graphicspath{{Figures/}} %
\begin{document}

\title{Spectroscopic Properties of $\mathit {B}$ and $\mathit {B_s}$ meson using Screened Potential}

\author{ Vikas Patel$^{a,b}$, Raghav Chaturvedi$^c$, A. K. Rai$^a$
}                     % Do not remove
\offprints{patelvikas2710@gmail.com\\
raghavr.chaturvedi@gmail.com\\
raiajayk@gmail.com}          % Insert a name or remove this line
\institute{$^{a}$Department of Applied Physics, Sardar Vallabhbhai National Institute of Technology, Surat-395007, Gujarat, {\it INDIA}\\
$^{b}$Department of Physics, Uka Tarsadia University, Bardoli 394250, Gujarat, {\it INDIA} \\
$^{c}$ Ministry of Education, Dubai, {\it U.A.E}}%
\date{Received: date / Revised version: date}
\abstract{
Inspired by the recent observations of $B$ and $B_s$ meson states at LHCb\cite{Aaij:2020hcw} we calculate the masses of ground, orbitally and radially excited states. Also, we estimate the mixing parameters, decay constant, leptonic decay width, and corresponding branching ratios, as well as electromagnetic transition widths. ${\cal{O}}\left(\frac{1}{m}\right)$ and ${\cal{O}}(p^{10})$ relativistic corrections to potential and kinetic energy terms have been added to the Hamiltonian. The screening potential employed is solved by applying the gaussian wave function. The estimated masses are used to constructed the Regge trajectories, which help us in association of some newly observed states to $B$ and $B_s$ meson family. Overall, the results from present study are in fair agreement with available experimental and theoretical studies.
\PACS{
      {PACS-key}{discribing text of that key}   \and
      {PACS-key}{discribing text of that key}
     } % end of PACS codes
} %end of abstract

\authorrunning { V Patel, R Chaturvedi, A K Rai}
\titlerunning {Spectroscopic Properties of $\mathit {B}$ and $\mathit {B_{s}}$ meson using Screened Potential}
\maketitle

\section{Introduction}
\label{sec1}
Properties of first two observed states of B-meson, $B^0$ and $B^\pm$ are well established since their discovery in 1983 at CLEO\cite{PhysRevLett.50.881}. After two years lowest-lying vector B-meson ($B^*$) of $J^P = 1^-$ was observed\cite{PhysRevLett.55.36}. The mass difference between ($m_{B^{*}}- m_{B}$) is the best measured quantity and its value is $45.18 \pm 0.23$. $B_{J}^{*}(5732)$ is the first orbitally excited B-meson was observed at OPAL detector at LEP \cite{Akers:1994fz}. Later in year 2000 it was studied again at OPAL detector in the ${B^{*}}{\pi}$ invariant mass distribution and the mass was determined to be $5738\pm 7$ Mev\cite{Abbiendi:2000zv}. In the year 2007, two orbitally excited (L=1) narrow B-mesons were observed at $D\Phi$ in ${B^{*}+}{\pi}^{-}$ invariant mass distribution\cite{Abazov:2007vq}. They are ${{ B}_{{1}}{(5721)}^{0}}$ with $J^P = 1^+$ and ${{ B}_{{2}}^{*}{(5747)}^{0}}$ with $J^P = 2^+$. In 2013 $B(5970)$ resonance with masses $5978 \pm 5 \pm 12$ Mev for neutral state and $5961\pm 5 \pm 12$ Mev for the charged state was observed at CDF collaboration in $B^{+}\pi^{-}$ and $B^{0}\pi^{+}$ mass distribution\cite{Aaltonen:2013atp}. In the year 2015 LHCb collaboration\cite{Aaij:2015qla} discovered another excited B-meson $B_{J}(5960)$. But studies at CDF\cite{Aaltonen:2013atp} suggested that $B_{J}(5960)$ resonances are the same resonant states of $B(5960)$. In addition to $B_{J}(5970)$ in 2015, the LHCb collaboration\cite{Aaij:2015qla} also observed another resonant state $B_{J}(5840)$ with masses $5869 \pm 5 \pm 6.7 \pm 0.2$ Mev for neutral state $B_{J}(5840)^{0}$ and $5850.3 \pm 12.7 \pm 13.7 \pm 0.2 $ Mev for charged state $B_{J}(5840)^{+}$.\\
In the $B_{s}$-meson family the $B_s$ and $B_s^*$ were observed in year 1990 at CUSB-II\cite{LeeFranzini:1990gy}, with masses $5366.82 \pm 0.22$ Mev and $5415.4 ^{+1.8}_{-1.5} $ Mev respectively. The first excited $B_{s}$-meson was observed in year 1994 as $B_{sJ}^{*}(5850)$ at LEP in $B^{(*)+}K^{-}$ invariant mass distribution\cite{Akers:1994fz} with mass $5853\pm 15$ Mev. Later, at LHCb the first orbitally excited $B_{s}$-meson states, ${{ B}_{{s1}}{(5830)}}$ and ${{ B}_{{s1}}{(5840)}}$ with masses $5828.63 \pm 0.27$ Mev and $5839.85 \pm 0.19$ Mev were observed at CDF collaboration in 2007\cite{Aaltonen:2007ah}. Later, these two states were confirmed by D$\Phi$ and LHCb\cite{Aaij:2012uva,Abazov:2007af}. Recently, LHCb collaboration \cite{Aaij:2020hcw} observed two new $B_s$ meson states $B_{sJ}(6064)$ and $B_{sJ}(6114)$. In future at LHCb more and more excited $B$ and $B_s$ meson states are expected to be observed due to its vast production cross-section of beauty, together with a good reconstruction efficiency, versatile trigger scheme and an excellent momentum and mass resolution\cite{Belyaev:2021cyr}. Listed in Table[\ref{Table:intro table}] are the various $B$ and $B_s$ mesons with experimental states, masses and experimental facilities where they were first observed.\\
With increased number of observed new experimental states many theoretical studies have attempted to study $B$ and $B_s$ mesons. The mass spectroscopy has been calculated by various models like relativistic or non-relativistic quark models, potential model, Bethe-Salpeter equation as well as constituent quark model\cite{Abazov:2007vq,Aaltonen:2008aa,Aaltonen:2007ah,Aaij:2015qla,Belyaev:2021cyr,Cheng:2017oqh,Kher:2017mky,Ebert:2009ua,Chen:2016spr,Chen:2018nnr}. Also, the strong decays, radiative decays, and semileptonic decays have been calculated theoreticaly\cite{Sun:2014wea,Godfrey:2016nwn,Lu:2016bbk,Asghar:2018tha,Yu:2019sqp,Aliev:2018kry,Gan:2010hw}.\\
$B_{1}(5721)$, $B_{2}^{*}(5747)$, $B_{S1}(5830)$, $B_{S2}^{*}(5840)$ are well entrenched states and various quark model studies \cite{Gan:2010hw,Liu:2016efm,Alhakami:2020vil,Sun:2014wea,Godfrey:2016nwn,Lu:2016bbk,Asghar:2018tha,Godfrey:1985xj,Zeng:1994vj,DiPierro:2001dwf} have tried to prove them as first excitations of $B$ and $B_{s}$ meson. Since, their discovery at CDF and LHCb, based on their observed masses and decay properties, theoretical studies have associated $B_{J}(5840)$ as $B(2^1 S_0)$ \cite{Lu:2016bbk,Asghar:2018tha,Gupta:2017bcm,li:2021hss}, $B(2^3 S_1)$ by \cite{Yu:2019iwm}, or $B(1^3 D_1)$ by \cite{Gupta:2017bcm}. Whereas, $B_{J}(5960)^{0}$ is associated as $B(2^3 S_1)$ by \cite{Liu:2016efm,Sun:2014wea,Gupta:2017bcm}, $B( 1^3 D_3)$ by \cite{Lu:2016bbk,Yu:2019iwm}, or $B(1^3 D_1)$ by \cite{Asghar:2018tha,Xiao:2014ura}. Their is only one theoretical study about newly observed $B_s$ meson states, $B_{sJ}(6064)$ and $B_{sJ}(6114)$. Thus, providing us inspiration to study the $B$ and $B_s$ meson spectroscopy. In this article we employ screening potential to study the mass spectroscopy and various decay properties of $B$ and $B_s$ mesons. We expect that for the light quark confining term will play a crucial role along with usual Coulombic term. Also, the screening potential will help to study the effective quenching between the quark and anti-quark. We then solve the potential using the variational approach. We also include ${\cal{O}}(p^{10})$ relativistic correction to kinetic energy and potential energy terms in the Hamiltonian. The Regge trajectories are constructed to help the association of new experimental observed states to the $B$ and $B_s$ meson family. We also estimate the pseudoscalar and vector decay constants as these decay constants are very important for calculating various weak decay properties. The leptonic branching ratio (BR) and electromagnetic (EM)transitions are also calculated in this scheme.\\
After introduction, Section \ref{sec:Theoretical framework} deals with theoretical framework to calculate the mass spectroscopy. Section \ref{sec:Regge} contains Regge trajectories. Section \ref{sec:DecayConstant} accomodates calculations for decay constant. In Section \ref{sec:leptonic branching fractions} the leptonic branching fractions are computed. Section \ref{sec:mix} deals with mixing parameters' calculations. In Section \ref{sec:EM}, calculation of electromagnetic transition widths are discussed and finally, in Section \ref{sec:Result} results, discussion and conclusion are presented.

\begin{table*}
%tiny
%\scriptsize
\caption{\label{Table:intro table} $B$ and $B_{s}$ mesons with experimental states, masses and experimental facilities where they were first observed.}
\centering
\vspace{0.1cm}
\begin{tabular}{ccccc}
\hline
 State & $J^P$ & Mass (MeV) &  Observed Mode &Experiments
\\
\hline
$B^{\pm}$ & $0^-$ & $5279.31 \pm 0.15 $ & $D\pi\&D^{*}\pi\pi$ & CLEO\cite{behrends1983observation}  \\
$B^{0}_{s}$ & $0^-$ & $5366.82 \pm 0.22$ &  & CUSB-II\cite{lee1990hyperfine}
\\
$B^{0}$ &  $0^-$ & $5279.58 \pm 0.15 \pm 0.28$ &$D\pi\pi\&D^{*}\pi$  & CLEO\cite{behrends1983observation}
\\ \hline
$B^{*}$ & $1^{-}$ & $5324.65 \pm 0.25$ &$ B \gamma $  & CUSB\cite{han1985observation}  \\
$B^{*}_{s}$ & $1^-$ & $5415.4 ^{+1.8}_{-1.5} $ & $B_{s} \gamma $ & CUSB-II\cite{lee1990hyperfine}
\\
$B_{1}(5721)^{*\pm}$ & $1^{+}$ & $5725^{+2.5}_{-2.7}$ &$B ^{* 0}\Pi ^{-} $& LHCb\cite{aaij2015precise} \\

$B_{1}(5721)^{0}$ & $1^{+}$ & $5726\pm 1.3$ &$B ^{* +}\Pi ^{-} $& CDF\cite{aaltonen2014study} \\
 $B_{s1}(5830)$ & $1^{+}$ & $5828.40 \pm 0.04\pm 0.04\pm 0.41$ & $B^{*}K$& LHCb\cite{aaij2013first}
\\

$B^{*}_{s2}(5840)^{0}$ & $1^{+}$ & $5839.99 \pm 0.05\pm 0.11\pm 0.17$ & $B^{*}K$& LHCb\cite{aaij2013first}
\\

 \hline
$B_{J}(5732)$ & $?^?$ & $5695^{+17}_{_19}$ &$B ^{*}\Pi  $& ALEPH\cite{barate1998resonant} \\

$B^{*}_{2}(5747)^{+}$ & $2^{+}$ & $5737.20\pm0.72\pm0.40\pm0.17$ &$B ^{*0}\Pi^{+}$& LHCb\cite{aaij2015precise}\\

  $B^{*}_{sJ}(5850)$ & $?$ & $5853 \pm 15$ & & OPAL\cite{opal1994observations}
\\

$B^{*}_{2}(5747)^{0}$ & $2^{+}$ & $5739.44\pm 0.37\pm0.33\pm 0.17$ &$B ^{*+}\Pi^{+}$&  LHCb\cite{aaij2015precise}\\

$B_{J}(5840)^{+}$ & $?^?$ & $5850.3\pm 12.7\pm 13.7\pm 0.2$ &$B\Pi$ &LHCb\cite{aaij2015precise} \\

$B_{J}(5840)^{0}$ & $?^?$ & $5862.9\pm 5.0\pm 6.7\pm 0.2$ &$B \Pi$&LHCb\cite{aaij2015precise} \\

$B_{J}(5970)^{+}$ & $?^?$ & $5961\pm 5 \pm 12$ &$B ^{0}\Pi^{+}$&CDF\cite{aaltonen2014study} \\

$B_{J}(5970)^{0}$ & $?^?$ & $5978\pm 5\pm 12$ &$B ^{+}\Pi^{-}$&CDF\cite{aaltonen2014study} \\
\hline
$B_{sJ}(6064)^{0}$ & $?^?$ & $6063.5\pm 1.2 \pm 0.8$ & $B ^{+}K^{-}$ & LHCb\cite{Aaij:2020hcw}\\
$or B_{sJ}(6109)^{0}$ & $?^?$ & $6108.8\pm 1.1 \pm 0.7$ & $B^{*+}K^{-}$ & LHCb\cite{Aaij:2020hcw}\\
\hline
$B_{sJ}(6114)^{0}$ & $?^?$ & $6114\pm 3 \pm 5$ & $B ^{+}K^{-}$ & LHCb\cite{Aaij:2020hcw} \\
$or B_{sJ}(6158)^{0}$ & $?^?$ & $6158.5\pm 4 \pm 5$ & $B ^{*+}K^{-}$ & LHCb\cite{Aaij:2020hcw}\\
\hline
\end{tabular}
\end{table*}

\section{Theoretical framework for Mass Spectroscopy calculation}
\label{sec:Theoretical framework}
We use the below mentioned Hamiltonian to calculate the $B$ and $B_s$ meson mass spectrum\cite{PhysRevD.84.074030},
\begin{equation}
\it{H=\sqrt{{p}^{2}+m_{1}^{2}}+\sqrt{{p}^{2}+m_{\bar{2}}^{2}}+V(\mathbf{r});\label{Eq:hamiltonian}}
\end{equation}
$m_1$ and $m_{\bar{m_2}}$ are the heavy$(b, \bar{b})$ and light quark$(u, \bar{u}, d, \bar{d}, s, \bar{s})$ masses, $p$ is the relative momentum of the quark and anti-quark system, and $V(\mathbf{r})$ is the quark anti-quark interaction potential. To accompany the relativistic effects the kinetic energy term in the Hamiltonian has been expanded to ${\cal{O}}(p^{10})$. In Table \ref{tab:p orders} we list down the contribution to the Hamiltonian due to various orders in $p$ (${\cal{O}}(p)$). Except ${\cal{O}}(p^{2})$ the leading terms in ${\cal{O}}(p)$ have shrinking values, but because ${\cal{O}}(p^{4})$ and ${\cal{O}}(p^{8})$ terms are negative and ${\cal{O}}(p^{6})$ and ${\cal{O}}(p^{10})$ terms are positive, they may have cancelling effect on each other, as a consequence we feel the term ${\cal{O}}(p^{10})$ may play a significant role in the mass spectroscopy calculation.
The quark anti-quark potential is of the form\cite{PhysRevLett.97.122003},
\begin{equation}
\it{V\left(r\right)=V^{\left(0\right)}\left(r\right)+\left(\frac{1}{m_{1}}+\frac{1}{m_{\bar{2}}}\right)V^{\left(1\right)}\left(r\right)+{\cal O}\left(\frac{1}{m^{2}}\right);\label{Eq:potential}}
\end{equation}
Where, $V^{\left(0\right)}$ is the spin-independent potential between the quark and anti-quark\cite{RevModPhys.80.1161}, $V_v(r)$ is Lorentz vector and $V_s(r)$ is Lorentz scalar contribution.
\begin{eqnarray}
\it{V^{\left(0\right)}\left(r\right)=V_v(r)+ V_s(r) + V_{0};\label{Eq:eqation01}}
\end{eqnarray}
\begin{eqnarray}
V_v(r)=-\frac{\alpha_c}{r}
\end{eqnarray}
\begin{equation}
 \it{ V_s(r) \equiv
\begin{cases}
Ar & linear \\
\frac{A}{\mu}(1-e^{-\mu r})  &screened ;\label{Eq:eqation screen}
\end{cases}}
\end{equation}
It has been interpreted by theoretical studies\cite{Laermann:1986pu,Born:1989iv}, at distances greater than 1 $fm$ the spontaneous creation of light quark anti-quark pairs inside a meson softens linear confinement potential by screening a colour charge. Hence, we research the effect of screening on linear confinement potential at larger distances in the quark anti-quark interaction. In present article we modify the Cornell potential by incorporating the Screening effect and calculate the ground and excited state masses for $B$ and $B_s$ mesons. At distance $r\ll 1/\mu$, $V_s(r)$ behave like linear potential ($A(r)$) and for $r\gg 1/\mu$, $V_s(r)$ becomes constant. Where, ``$\mu$'' is the screening factor and its value during the mass spectra calculation is taken as $0.04$, and $V_{0}$ is potential constant. To justify why the value of ``$\mu$'' is taken as $0.04$ during the mass spectra calculation for $B$ and $B_s$ meson, masses for various states of have been calculated for different values of ``$\mu$''and are tabulated in Tables \ref{tab:mu} and \ref{tab:mu1}, the calculated masses from the present work matches well with the experimental masses for $\mu = 0.04$. Also, the $\chi^2/\textit{d.o.f}$ or the goodness of fit values for different ``$\mu$'' values for $B$ and $B_s$ mesons have been calculated. The $\chi^2/\textit{d.o.f}$ is calculated as per \cite{Chaturvedi_2020}, $\chi^2/\textit{d.o.f}= \sum \frac{(observed-expected)^2}{expected}$. Here, observed values are the predicted values for various $\mu's$ and expected values are the experimental masses, Minimum value for  $\chi^2/\textit{d.o.f}$ is obtained for $\mu = 0.04$. Hence, validating the choice of ``$\mu$'' as $0.04$. $\alpha_c$ and $\alpha_s$ are effective running and running coupling constants determined through the simplest model with freezing\cite{PhysRevD.70.016007,Simonov:1993kt,PhysRevD.79.114029}. $\alpha_c$ and $\alpha_s$ follow the relation  $\alpha_c = 4/3 \alpha_s$. $\Lambda$, the QCD scales taken as 0.413 GeV. $n_f$ is the number of flavors and $M$ is the re-normalisation scale related to the constituent quark masses as $M = 2m_Q m_{\bar{q}}/(m_Q + m_{\bar{q}})$\cite{PhysRevD.79.114029}. $A$ in Eq. \ref{Eq:eqation screen} represents potential strength.
\begin{eqnarray}
     \alpha_{s}(\mu)^2=\frac{4\pi}{\left(11-\frac{2}{3}n_{f}\right)\ln\frac{M^{2}+{M_B}^2} {\Lambda^{2}}}
\end{eqnarray}
$V^{\left(1\right)}\left(r\right)$ in Eq. \ref{Eq:potential} is a relativistic correction taken as,
\begin{equation}
V^{\left(1\right)}\left(r\right)=-C_{1}C_{2}\alpha_{s}^{2}/4r^{2}
\end{equation}
$C_{1}=4/3$ and $C_{2}=3$ are the Casimir charges. The values of all the potential parameters used in the present work are listed in Table \ref{Table:Parameter}.
In the present work Gaussian like wave-function in position and momentum space has been used, see Eqs. \ref{Eq:Gauss1} \& \ref{Gauss2}. Where, ``$u$'' and ``$L$'' are variational parameter and Laguerre polynomial, respectively. Ritz variational scheme has been used. And, the expectation value is obtained as $ H \psi = E \psi$.
\begin{eqnarray}
\it {R_{nl}(u,r)} & = & \it {u^{3/2}\left(\frac{2\left(n-1\right)!}{\Gamma\left(n+l+1/2\right)}\right)^{1/2}\left(u r\right)^{l}\times e^{-u^{2}r^{2}/2}L_{n-1}^{l+1/2}(u^{2}r^{2});\label{Eq:Gauss1}}
\end{eqnarray}
  \cal\begin{eqnarray}
\it {R_{nl}(u,p)} & = & \it {\frac{\left(-1\right)^{n}}{u^{3/2}}\left(\frac{2\left(n-1\right)!}{\Gamma\left(n+l+1/2\right)}\right)^{1/2}\left(\frac{p}{u}\right)^{l} e^{-{p}^{2}/2u^{2}}L_{n-1}^{l+1/2}\left(\frac{p^{2}}{u^{2}}\right);\label{Gauss2}}
\end{eqnarray}
The values of the variational parameter ``$u$'' has been found graphically using the virial theorem\cite{PhysRevD.55.6944}. The values for ``$u$'' can be found in Table \ref{tab:SA}. The Gaussian wave function in the position space is employed to obtain the expectation value of potential energy. Whereas, the Gaussian wave function in the momentum space is employed to obtain the expectation value of kinetic energy.
\begin{equation}
 \left\langle{K.E.}\right\rangle =\frac{1}{2} \left\langle{\frac{rdV}{dr}}\right\rangle.
\end{equation}
Equation \ref{Eq:rai1-1} is employed to determine the spin-average mass for the ground state\cite{PhysRevC.78.055202}. The potential constant $V_{0}$ in Eq. \ref{Eq:eqation01} is fixed to obtain the experimental spin-averaged mass. The expectation value of the Hamiltonian yield spin-average mass. The calculated spin-average masses can be found in Table \ref{tab:SA}.
\begin{equation}
M_{SA}=M_{PS}+\frac{3}{4}(M_{Vec}-M_{PS})\label{Eq:rai1-1}
\end{equation}
We compute the spin-average mass from the  respective theoretical values as per Eq.\ref{Eq:rai2-1} in Ref.\cite{PhysRevC.78.055202} for the comparison for the $nJ$ state,
\begin{equation}
M_{CW,n}=\frac{\Sigma_{J}(2J+1)M_{nJ}}{\Sigma_{J}(2J+1)}\label{Eq:rai2-1}
\end{equation}
where, $M_{CW,n}$ and $M_{nJ}$ denotes the spin-averaged mass of the $n$ state, and the mass of the meson in the $nJ$ state, respectively.
\begin{equation}
M_{CW,n}=\frac{\Sigma_{J}(2J+1)M_{nJ}}{\Sigma_{J}(2J+1)}\label{Eq:rai2-1}
\end{equation}
We add perturbatively the spin-dependent potential to calculate the hyperfine and spin-orbit splitting which is of the form\cite{PhysRevD.49.5845},
\begin{eqnarray}
V_{spin}(r) & = & \left(\frac{{L\cdot S_{1}}}{2m_{1}^{2}}+\frac{{L\cdot S_{\bar{2}}}}{2m_{\bar{2}}^{2}}\right)\left(-\frac{dV^{\left(0\right)}(r)}{rdr}+\frac{8}{3}\alpha_{S}\frac{1}{r^{3}}\right)+\nonumber \\
 &  & \frac{4}{3}\alpha_{S}\frac{1}{m_{1}m_{\bar{2}}}\frac{{L\cdot S}}{r^{3}}+\frac{4}{3}\alpha_{S}\frac{2}{3m_{1}m_{\bar{2}}}{S_{1}\cdot S_{\bar{2}}}4\pi\delta({r})\label{eq:spinhyperfine}\nonumber\\
 &  & +\frac{4}{3}\alpha_{S}\frac{1}{m_{1}m_{\bar{2}}}\Biggl\{3({S_{1}\cdot n})({S_{\bar{2}}\cdot n})-({S_{1}\cdot S_{\bar{2}}})\Biggr\}\frac{1}{r^{3}},\ \quad{n}=\frac{{r}}{r} ;\label{Eq:spinhyperfine}
\end{eqnarray}
The first term in Eq. \ref{Eq:spinhyperfine} is the relativistic correction to the potential $V^{\left(0\right)}$, the second term considers the spin-orbit interaction, the third and the fourth terms are for spin-spin and tensor interactions. The heavy-heavy flavored meson states with $J=L$ are mixtures of spin-triplet $\left|^{3}L_{L}\right>$ and spin-singlet $\left|^{1}L_{L}\right>$
states: $J=L=1,\ 2,\ 3,\ldots$

\begin{eqnarray}
\left|\psi_{J}\right> & = & \left|^{1}L_{L}\right>\cos{\theta}+\left|^{3}L_{L}\right>\sin{\theta}
\end{eqnarray}
\begin{eqnarray}
\left|\psi_{J}^{\prime}\right> & = & -\left|^{1}L_{L}\right>\sin{\theta}+\left|^{3}L_{L}\right>\cos{\theta} \end{eqnarray}
\noindent where, ``$\theta$'' is the mixing angle and the primed state has the heavier mass. Such mixing occurs due to the nondiagonal spin-orbit and tensor terms in Eq.\ref{Eq:spinhyperfine}. The calculated masses can be found in Tables \ref{tab:Bmass}, \ref{tab:Bmass1}, \ref{tab:Bsmass}  \& \ref{tab:Bsmass1}.

\begin{table*}[!htb]
%\tiny
%\scriptsize
\caption{Potential Parameters}. \\
\centering
\resizebox{\textwidth}{!}{
\begin{tabular}{ccccccccccc}
%\hline\noalign{\smallskip}
\hline\noalign{\smallskip}
Meson&$ \alpha_{s} $  &$ \alpha_{c} $ &$m_{u/d}$ &$m_{s}$ & $m_{b}$ &nf  & $ \Lambda $ & $ \mu $& A &$ V_{0} $\\
&&&(GeV)&(GeV)&(GeV)&&&&$(GeV^2)$&\\
\hline
$B$&0.732&0.976&0.46&&4.53&5&0.413&0.04&0.16&-0.095\\
$B_{s}$&0.668&0.891&&0.586&4.53&5&0.413&0.04&0.16&0.004\\

\hline\noalign{\smallskip}
\end{tabular}
}
$\label{Table:Parameter}$
\end{table*}
{\cal \begin{table*}
\begin{center}

\caption{\label{tab:p orders} 1S, 2S \& 1P spin-average masses of $B$ and $B_s$ mesons considering various orders of ${\cal{O}}(p)$.}
\begin{tabular}{|c|cccc|cccc|}
\hline
&&&$B$-meson&&&&$B_s$-meson&\\
\hline
 state & ${\cal{O}}(p^{4})$ & ${\cal{O}}(p^{6})$ & ${\cal{O}}(p^{8})$ & ${\cal{O}}(p^{10})$& ${\cal{O}}(p^{4})$ & ${\cal{O}}(p^{6})$ & ${\cal{O}}(p^{8})$ & ${\cal{O}}(p^{10})$\\
\hline
1S&4667	&4808	&4523	&5314&4800	&5949	&5667	&5401\\
\hline
2S&5448	&5626	&5326	&5959&5567	&5741	&5475	&5990\\
\hline
1P&5455	&5571	&5385	&5779&5596	&5701	&5550	&5838\\
\hline
1D&5794	&5922	&5729	&6104&5923	&6033	&5886	&6139\\

\hline
\end{tabular}
\end{center}
\end{table*}}

{\cal \begin{table*}
\begin{center}
\caption{\label{tab:mu} Masses for $B$ meson for various $\mu$ from $0.01$ till $0.05$.}
\begin{tabular}{ccccccc}
\hline
 state  &  0.01 & 0.02 &  0.03 &  0.04 &  0.05  \\
\hline
$1^1S_{0}$ &  5378&5322&5290 &5282 &5275 \\
$1^3S_{1}$  & 5364&5364&5331 &5324 &5317 \\
$2^1S_{0}$ &   6163&6069&5981 & 5951&5923 \\
$2^3S_{1}$  & 6174&6080&5991 & 5962&5934 \\
$3^1S_{0}$ &  6832&6589&6511 & 6449&6391 \\
$3^3S_{1}$ &     6837&6593&6516 & 6453&6396  \\
\hline
$1^3P_{0}$ &  5810&5790&5771 & 5752&5735 \\
$1^3P_{1}$  &  5822&5803&5784 & 5766&5748 \\
$1^1P_{1}$& 5837&5818&5800 & 5783&5766 \\
$1^3P_{2}$ &   5835&5815&5797 & 5779&5762 \\
$2^3P_{0}$  &  6442&6389&6338 & 6291&6247\\
$2^3P_{1}$  &   6449&6396&6346 & 6299&6255\\
$2^1P_{1}$ &   6457&6404&6355 & 6309&6247\\
$2^3P_{2}$ &   6456&6403&6354 & 6307&6263\\
\hline
$1^3D_{1}$ &  6210&6176&6143 & 6112&6083\\
$1^3D_{2}$ & 6199&6164&6134 & 6104&6075\\
$1^1D_{2}$ &  6198&6164&6132 & 6102&6074\\
$1^3D_{3}$ & 6195&6163&6132 & 6102&6073\\
$2^3D_{1}$ &  6787&6711&6642 & 6577&6517\\
$2^3D_{2}$  &6778&6703&6635 & 6571&6511\\
$2^1D_{2}$&6778&6703&6634 & 6570&6510\\
$2^3D_{3}$  &6774&6700&6632 & 6569&6510\\
\hline
$1^3F_{2}$ &6527&6746&6428 & 6383&6340\\
$1^3F_{3}$ &6515&6765&6418 & 6374&6332\\
$1^1F_{3}$ &6513&6764&6417 & 6373&6331\\
$1^3F_{4}$ &6505&6457&6411 & 6368&6327\\
\hline
\end{tabular}
\end{center}
\end{table*}}

{\cal \begin{table*}
\begin{center}
\caption{\label{tab:mu1} Masses for $B_s$ meson for various $\mu$ from $0.01$ till $0.05$.}
\begin{tabular}{ccccccc}
\hline
 state  &  0.01 & 0.02 &  0.03 &  0.04 &  0.05  \\
\hline
$1^1S_{0}$ &  5374&5369&5364 &5359 &5354 \\
$1^3S_{1}$  & 5329&5424&5420 &5415 &5410 \\
$2^1S_{0}$ &   6044&6021&6000 & 5980&5960 \\
$2^3S_{1}$  & 6057&6035&6014 & 5993&5973 \\
$3^1S_{0}$ &  6580&6530&6483 & 6438&6397 \\
$3^3S_{1}$ &     6586&6536&6488& 6444&6403  \\
\hline
$1^3P_{0}$ &  5838&5824&5831 & 5798&5785 \\
$1^3P_{1}$  &  5858&5844&5811 & 5818&5806 \\
$1^1P_{1}$& 5884&5871&5828 & 5846&5834 \\
$1^3P_{2}$ &   5877&5864&5851 & 58389&5826 \\
$2^3P_{0}$  &  6399&6361&6326 & 6292&6260\\
$2^3P_{1}$  &   6410&6373&6338 & 6304&6272\\
$2^1P_{1}$ &   6425&6388&6323 & 6320&6289\\
$2^3P_{2}$ &   6422&6350&6349 & 6316&6284\\
\hline
$1^3D_{1}$ &  6203&6179&6157 & 6144&6144\\
$1^3D_{2}$ & 6205&6190&6167 & 6139&6123\\
$1^1D_{2}$ &  6214&6182&6160 & 6135&6119\\
$1^3D_{3}$ & 6206&6183&6160 & 6139&6118\\
$2^3D_{1}$ &  6706&6653&6603 & 6564&6513\\
$2^3D_{2}$  &6718&6661&6611 & 6560&6520\\
$2^1D_{2}$&6706&6654&6605 & 6557&6516\\
$2^3D_{3}$  &6707&66655&6606 & 6559&6516\\
\hline
$1^3F_{2}$ &6496&6460&6425 & 6392&6361\\
$1^3F_{3}$ &6486&6450&6416 & 6384&6353\\
$1^1F_{3}$ &6486&6450&6417 & 6385&6354\\
$1^3F_{4}$ &6479&6445&6412 & 6381&6351\\
\hline
\end{tabular}
\end{center}
\end{table*}}

\begin{table}
\begin{center}
{{\caption{\label{tab:SA}{Spin average masses for S, P, D and F states in $B$ and $B_{s}$ mesons (in MeV).}}}}
\resizebox{\textwidth}{!}{
\begin{tabular}{c c c c c c c c c c c c c}
\hline
Meson & State & $ \mu $(MeV) & M$_{SA}$(MeV) & \cite{tanabashi2018particle} & \cite{Kher:2017mky}   & \cite{shah2016spectroscopy} & \cite{Ebert:2009ua}  & \cite{Liu:2016efm}\\
\hline
 $B$ & 1S & 390  & 5314 & 5314 &5314  &5314 &5314&  5288  \\
   & 2S & 281  & 5951 &5944  &5942  &5819 &5902&  5903 \\
   & 3S & 231  &6452  &  &6504  &6251  &6385 \\
   & 4S &200  &6846  &  &6772   &  &6785 \\
   & 5S & 179   &7165  & &7546 &  &  7132  \\
   & 6S & 163 &7427  &  &  \\

   \hline
   & 1P& 314  & 5779 & 5730& 5740  &5737 &5745&  5759 \\
   & 2P &247  & 6307 & & 6301   &6127 &6249&  6188 \\
   & 3P &211  &6726  &&6828&6482 &6669\\
   & 4P & 187 &7068  &&\\
   & 5P & 169 &7348  &&\\

   \hline
   & 1D & 280  & 6104 & & 6057  &6045 &6106&   6042 \\
   & 2D & 229  & 6571 & & 6596  &6429 &6540&   6377 \\
   & 3D &199 &6944 & &7110& 6769\\
   & 4D &178  &7249  & &\\
   & 5D & 162 &7500 & &\\

   \hline
   & 1F & 258 & 6343 &&&& \\
   & 2F & 215 & 6798 & &&  \\
   & 3F & 189 &7136  & &\\
   & 4F & 171 &7411  & &\\
   & 5F & 157 &7642  & &\\

    \hline
     $B_{s}$ & 1S & 481 & 5401 & 5401& 5401  &5403 &5404&  5370 \\
   & 2S & 342  & 5990 & & 6023  &5952 &5988&  5971 \\
   & 3S &280  &6443  &  &6570 &6425  &6473  \\
   & 4S &242  &6788  &  & 7083  &6863  &6878 \\
   & 5S &219   &  7117& &7575 &   \\
   & 6S & 200 &7376  &  & & \\

   \hline
   & 1P& 379 & 5838 & 5840& 5835   &5838 &5844&  5838 \\
   & 2P &294 & 6316 & & 6380 &6233 &6343&   6254\\
   & 3P &255  &6701  &&6889&6603 &6768&&\\
   & 4P &227  & 7023 &\\
   & 5P &206  &7295  &&\\

   \hline
   & 1D & 335  & 6139 & & 6150   &6181 &6200&   6117 \\
   & 2D & 275  & 6560 & & 6668& 6626 &6635&   6450 \\
    & 3D &240 &6905 & &7162& 6912\\
   & 4D & 216 &7358  & &\\
   & 5D & 198 &7444 & &\\
   \hline
   & 1F & 307 & 6385 &&&&&\\
   & 2F & 259 &6769  & &&  \\
    & 3F & 229 & 7085 & &\\
   & 4F & 208 & 7352 & &\\
   & 5F & 191 & 7579 & &\\

    \hline
\end{tabular}
}
\end{center}
\end{table}

\begin{table}
\begin{center}
%\caption{{S and P states  mass spectrum  of  $B$  meson  (in GeV).}}
\caption{\label{tab:Bmass} S and P states  mass spectrum  of  $B$  meson  (in MeV).}
\resizebox{\textwidth}{!}{
\begin{tabular}{cccccccccccccc}
\hline
State & $J^P$ & Meson&Present &\cite{tanabashi2018particle} & \cite{Kher:2017mky}   & \cite{shah2016spectroscopy} & \cite{Ebert:2009ua} & \cite{Godfrey:2016nwn} & \cite{Liu:2016efm}\\
\hline
$1^1S_{0}$ & $0^{-}$  &$ B^{0} $& 5282 & 5279$\pm 0.015$ & 5287 & 5289 & 5280 & 5312& 5273\\
&&$ B^{\pm}$&&5279$\pm 0.014$\\
$2^1S_{0}$ & $0^{-}$ & $B_{J}(5840)^{0} $ &5951 & 5860$\pm 0.081$   & 5926 & 5804 & 5840 & 5904& 5893\\
$3^1S_{0}$ & $0^{-}$ &&6449 &  &6492 &6242 &6379  & 6335  \\
$4^1S_{0}$ &$0^{-}$&&6844 &  &7027  &6641  &6781  &6689&  \\
$5^1S_{0}$ &$0^{-}$& &7164 &  &7538 & &&6997  \\
 $6^1S_{0}$ &$0^{-}$& &7426 & \\

$1^3S_{1}$  & $1^{-}$ & $B^{*}$&5324 & 5324 $\pm 0.25$  & 5323 & 5325 & 5326 & 5371& 5331\\
$2^3S_{1}$  & $1^{-}$ &$B_{J}(5970)^{+} $  &5962 & 5967$\pm 5$ & 5947 & 5834 & 5906 & 5933&  5932\\
&&$B_{J}(5970)^{0} $& &5971$\pm 5$ &&\\
$3^3S_{1}$ & $1^{-}$ &&6453 &   &6508  &6254  &6387  &6355 \\
$4^3S_{1}$ &$1^{-}$& &6846 &   &7039  & 6649 &6786  &6703 \\
$5^3S_{1}$ &$1^{-}$& &7165 && 7549 & & &7008  \\
$6^3S_{1}$ &$1^{-}$& &7427 & & \\
\hline
$1^3P_{0}$ &$0^{+}$&  &5752&$5732^{\pm 0.005}_{\pm 0.02}$ & 5730  &5697 & 5749& 5756 & 5740 \\
$1^1P_{1}$ &$1^{+}$& $B_{1}(5721)^{0} $ &5766&5725$^{+2.5}_{-2.7} $ & 5733  &5723 & 5723& 5777 & 5815 \\
$1^3P_{1}$ & $1^{+}$ && 5783&5726$\pm 1.3$  & 5752   &5738 & 5774& 5784 & 5731 \\
$1^3P_{2}$ & $2^{+}$&$B^{*}_{2}(5747)^{+} $ & 5779&5737$\pm 0.7$ & 5740   &5754 & 5741& 5797 & 5746\\
&&$B^{*}_{2}(5747)^{0} $&&5739$\pm 0.7$ &&\\
$2^3P_{0}$ & $0^{+}$ & &6291& &6297  &6053 &6221 & 6213&  6188\\
$2^1P_{1}$ & $1^{+}$& &6299& &6295  &6106 & 6209& 6197 &6168  \\
$2^3P_{1}$ &$1^{+}$& &6309& &6311 &6131 &6281 & 6228 & 6221 & \\
$2^3P_{2}$ & $2^{+}$ & &6307& &6299 &6153 &6260 & 6213& 6179 \\
$3^3P_{0}$ & $0^{+}$ &&6715 &  &6826&6375&6629&6576\\
$3^1P_{1}$ & $1^{+}$ && 6721 & &6829 &6453&6650&6557\\
$3^3P_{1}$ & $1^{+}$& &6727& &6837  &6486&6685&6585\\
$3^3P_{2}$ & $2^{+}$ &&6726 &&6826  &6518&6678&6570\\
$4^3P_{0}$ & $0^{+}$ && 7059 &&&&&6890&\\
$4^1P_{1}$ & $1^{+}$ & &7063 &&&&&6872&\\
$4^3P_{1}$ & $1^{+}$ &&7068  &&&&&6897&\\
$4^3P_{2}$ & $2^{+}$ &&7068 &&&&&6883&\\

$5^3P_{0}$ & $0^{+}$ &&7341  &&&&&&\\
$5^1P_{1}$ & $1^{+}$& &7344 &&&&&&\\
$5^3P_{1}$ & $1^{+}$&&  7348&&&&&&\\
$5^3P_{2}$ & $2^{+}$& &7348&&&&&&\\ \hline
\end{tabular}
}
\end{center}
\end{table}

\begin{table}
\begin{center}
\caption{\label{tab:Bmass1} D and F states  mass spectrum  of  $B$  meson  (in MeV).}
\resizebox{\textwidth}{!}{
\begin{tabular}{ccccccccccccc}
\hline
State & $J^P$ & Meson&Present &\cite{tanabashi2018particle} & \cite{Kher:2017mky}   & \cite{shah2016spectroscopy} & \cite{Ebert:2009ua} & \cite{Godfrey:2016nwn} & \cite{Liu:2016efm}\\
  \hline
$1^3D_{1}$ &$1^{-}$&&6112& &6016  &6104 &6119 & 6110 & 6135  \\
$1^1D_{2}$ &$2^{-}$& & 6104& &6031 &6076 &6121 & 6095 & 5967  \\
$1^3D_{2}$ & $2^{-}$ & & 6102& &6065  &6065 &6103 & 6124& 6152  \\
$1^3D_{3}$ & $3^{-}$ & & 6102& &6085 &6041 &6091 &6106 & 5976  \\
$2^3D_{1}$ & $1^{-}$ & & 6577& &6562&6460 &6534 & 6475& 6445  \\
$2^1D_{2}$ & $2^{-}$& &  6571& &6575 &6440 &6554 & 6450& 6323  \\
$2^3D_{2}$ &$2^{-}$ & & 6570& &6602 &6429 &6528 & 6486 & 6456  \\
$2^3D_{3}$ & $3^{-}$ & & 6569& &6619  &6409 &6542 & 6460& 6329  \\
$3^3D_{1}$ & $1^{-}$ &&  6949&  &7081&6795&&6792\\
$3^1D_{2}$ & $2^{-}$ & & 6944 &  &7093&6768&&6767\\
$3^3D_{2}$ & $2^{-}$& & 6944  &  &7116&6779&&6800\\
$3^3D_{3}$ & $3^{-}$ & & 6943 &  &7130&6751&&6775\\
$4^3D_{1}$ & $1^{-}$ &&  7253 &&&&&&\\
$4^1D_{2}$ & $2^{-}$ &&  7249 &&&&&&\\
$4^3D_{2}$ & $2^{-}$ & & 7248 &&&&&&\\
$4^3D_{3}$ & $3^{-}$ && 7248 &&&&&&\\
$5^3D_{1}$ & $1^{-}$ &&  7503&&&&&&\\
$5^1D_{2}$ & $2^{-}$&&  7500&&&&&&\\
$5^3D_{2}$ & $2^{-}$& & 7500 &&&&&&\\
$5^3D_{3}$ & $3^{-}$&& 7499 &&&&&&\\ \hline
$1^3F_{2}$ &$2^{+}$&& 6383 &&&&&6387\\
$1^1F_{3}$ &$3^{+}$&&6374&&&&&6358\\
$1^3F_{3}$ & $3^{+}$&  &6373&&&&&6396\\
$1^3F_{4}$ & $4^{+}$ & & 6368&&&&&6364\\
$2^3F_{2}$ & $2^{+}$& & 6805&&&&&6704\\
$2^1F_{3}$ & $3^{+}$&&  6798&&&&&6673\\
$2^3F_{3}$ &$3^{+}$&& 6798&&&&&6711 \\
$2^3F_{4}$ & $4^{+}$ & & 6793&&&&&6679\\
$3^3F_{2}$ & $2^{+}$ &&  7141 &  &&&&&\\
$3^1F_{3}$ & $3^{+}$ && 7136 &  &&&&&\\
$3^3F_{3}$ & $3^{+}$&&  7136 &  &&&&&\\
$3^3F_{4}$ & $4^{+}$ && 7132 &  &&&&&\\
$4^3F_{2}$ & $2^{+}$ &&  7416&&&&&&\\
$4^1F_{3}$ & $3^{+}$ && 7411 &&&&&&\\
$4^3F_{3}$ & $3^{+}$&  & 7411&&&&&&\\
$4^3F_{4}$ & $4^{+}$ & & 7408&&&&&&\\
$5^3F_{2}$ & $2^{+}$ & & 7646 &&&&&&\\
$5^1F_{3}$ & $3^{+}$&&  7642 &&&&&&\\
$5^3F_{3}$ & $3^{+}$&& 7642  &&&&&&\\
$5^3F_{4}$ & $4^{+}$&& 7639 &&&&&&\\ \hline
\end{tabular}
}
\end{center}
\end{table}

\begin{table*}
\begin{center}
\caption{\label{tab:Bsmass} S and P states  mass spectrum  of  $B_s$  meson  (in MeV).}
\resizebox{\textwidth}{!}{
\begin{tabular}{cccccccccccccc}
\hline
State & $J^P$ & Meson&Present& \cite{tanabashi2018particle} & \cite{Kher:2017mky}   & \cite{shah2016spectroscopy} & \cite{Ebert:2009ua} & \cite{Godfrey:2016nwn} & \cite{Liu:2016efm}\\
\\
\hline
$1^1S_{0}$ & $0^{-}$  & $B^{0}_{S}$& 5359 & 5366 $\pm 0.19$ & 5367 & 5366 & 5372 & 5394& 5355\\
$2^1S_{0}$ & $0^{-}$ & & 5980 &  & 6003 & 5939 & 5976 & 5984 & 5962\\
$3^1S_{0}$ & $0^{-}$ & &6438 &  &6556  &6419 &6467  & 6410&6415 \\
$4^1S_{0}$ &$0^{-}$& &6786 &  &7071 &6859  &6874  &6759  \\
$5^1S_{0}$ &$0^{-}$& &7116  &  &7565 & &&7063  \\
$6^1S_{0}$ &$0^{-}$&& 7375  & & \\

$1^3S_{1}$  & $1^{-}$ & $B^{*}_{S}$& 5415 & $5415^{+1.8}_{-1.5}$     & 5413 & 5415 & 5414 & 5440&   5416\\

$2^3S_{1}$  & $1^{-}$ & &5993 &  & 6029 & 5956 & 5992 & 6012& 5999\\

$3^3S_{1}$ & $1^{-}$ & &6444 &   & 6575&6427  &6475  &6429 \\

$4^3S_{1}$ &$1^{-}$&& 6789  &   &7087  &6864  &  6879&6773 \\

$5^3S_{1}$ &$1^{-}$&& 7118  &  &7579 & &&7076  \\

$6^3S_{2}$ &$1^{-}$&& 7376  & & \\
\hline
$1^3P_{0}$ &$0^{+}$&& 5798& & 5812  &5799 & 5833& 5831  & 5782 \\
$1^1P_{1}$ &$1^{+}$&$B_{S1}$ $(5830)^{0}$& 5818&5828 $\pm 0.27$  &  5828   &5819 & 5865& 5857  & 5833 \\
$1^3P_{1}$ & $1^{+}$ &$B^{*}_{sJ}$  &  5846&5850 & 5842   &5854 & 5831& 5861  & 5843 \\
$1^3P_{2}$ & $2^{+}$ &$B^{*}_{S2}$ $(5840)^{0}$ & 5838&5839 $\pm 0.17$ &  5840  &5849 & 5842& 5876& 5848\\
$2^3P_{0}$ & $0^{+}$&  & 6292& &6367   &6171 & 6348& 6279 &6220  \\
$2^1P_{1}$ & $1^{+}$& & 6304& &6375  &6197 & 6345& 6279&6250  \\
$2^3P_{1}$ &$1^{+}$ & &6320& &6387  &6278 &6321 & 6296& 6256 & \\
$2^3P_{2}$ & $2^{+}$ &&  6316& &6382  &6241 &6359 & 6295& 6261 &\\
$3^3P_{0}$ & $0^{+}$ & &6684 &  &6879&6510&6731&6639\\
$3^1P_{1}$ & $1^{+}$ & & 6693  &  &6885&6663&6761&6635\\
$3^3P_{1}$ & $1^{+}$&& 6704  &  &6895&6543&6768&6650\\
$3^3P_{2}$ & $2^{+}$ && 6701 &  &6890&6622&6780&6648\\
$4^3P_{0}$ & $0^{+}$ && 7009  &&&&&6950\\
$4^1P_{1}$ & $1^{+}$ &&  7016 &&&&&6946\\
$4^3P_{1}$ & $1^{+}$ &&  7025 &&&&&6959\\
$4^3P_{2}$ & $2^{+}$ && 7023 &&&&&6956\\
$5^3P_{0}$ & $0^{+}$ &&  7283 &&&&&&\\
$5^1P_{1}$ & $1^{+}$&&7289  &&&&&&\\
$5^3P_{1}$ & $1^{+}$&&  7297 &&&&&&\\
$5^3P_{2}$ & $2^{+}$&& 7295 &&&&&&\\
\hline
\end{tabular}
}
\end{center}
\end{table*}

\begin{table*}
\begin{center}
\caption{\label{tab:Bsmass1} D and F states  mass spectrum  of  $B_s$  meson  (in MeV).}
\resizebox{\textwidth}{!}{
\begin{tabular}{ccccccccccccc}
\hline
State & $J^P$ & Meson&Present& \cite{tanabashi2018particle} & \cite{Kher:2017mky}   & \cite{shah2016spectroscopy} & \cite{Ebert:2009ua} & \cite{Godfrey:2016nwn} & \cite{Liu:2016efm}\\
\\
\hline
$1^3D_{1}$ &$1^{-}$&  &6144& &6119  &6226 &6209 & 6179& 6155  \\
$1^1D_{2}$ &$2^{-}$&$B_{sJ}(6064) $ &6139& &6128 &6177 &6218 & 6169& 6079  \\
$1^3D_{2}$ & $2^{-}$ && 6135& &6157 &6209 &6189 & 6196& 6172  \\
$1^3D_{3}$ & $3^{-}$ & & 6139& &6172&6145 &6191 & 6179 & 6088  \\
$2^3D_{1}$ & $1^{-}$ && 6564& &6642 &6595 &6629 & 6542& 6478  \\
$2^1D_{2}$ & $2^{-}$&&  6560& &6650 &6554 &6651 & 6526& 6422 \\
$2^3D_{2}$ &$2^{-}$ & &6557& &6674 &6585 &6625 & 6553 & 6490  \\
$2^3D_{3}$ & $3^{-}$ && 6559& &6687 &6528 &6637 & 6535& 6429  \\
$3^3D_{1}$ & $1^{-}$ &&6909 &  &7139&6942&&6855\\
$3^1D_{2}$ & $2^{-}$ &&6905  &  &7147&6907&&6841\\
$3^3D_{2}$ & $2^{-}$& &6904 &  &7167&6936&&6864\\
$3^3D_{3}$ & $3^{-}$ &&6905  &  &7178&6885&&6849\\
$4^3D_{1}$ & $1^{-}$ &&7361  &&&&&&\\
$4^1D_{2}$ & $2^{-}$ && 7358 &&&&&&\\
$4^3D_{2}$ & $2^{-}$ &&7456  &&&&&&\\
$4^3D_{3}$ & $3^{-}$ &&7358 &&&&&&\\
$5^3D_{1}$ & $1^{-}$ &&7447 &&&&&&\\
$5^1D_{2}$ & $2^{-}$&&7444 &&&&&&\\
$5^3D_{2}$ & $2^{-}$&& 7443 &&&&&&\\
$5^3D_{3}$ & $3^{-}$&&7444 &&&&&&\\
\hline
$1^3F_{2}$ &$2^{+}$&&6392 &&&&&6454\\
$1^1F_{3}$ &$3^{+}$&&6384&&&&&6425\\
$1^3F_{3}$ & $3^{+}$ &&6385 &&&&&6462\\
$1^3F_{4}$ & $4^{+}$ & &6381&&&&&6432\\
$2^3F_{2}$ & $2^{+}$&& 6775&&&&&6768\\
$2^1F_{3}$ & $3^{+}$& &6769&&&&&6742\\
$2^3F_{3}$ &$3^{+}$&&6769&&&&&6775 \\
$2^3F_{4}$ & $4^{+}$ && 6765&&&&&6748\\
$3^3F_{2}$ & $2^{+}$ &&7090  &  &&&&&\\
$3^1F_{3}$ & $3^{+}$ &&  7085 &&&&&\\
$3^3F_{3}$ & $3^{+}$&& 7085 &  &&&&&\\
$3^3F_{4}$ & $4^{+}$ &&7082 &  &&&&&\\
$4^3F_{2}$ & $2^{+}$ &&7357 &&&&&&\\
$4^1F_{3}$ & $3^{+}$ &&7352 &&&&&&\\
$4^3F_{3}$ & $3^{+}$ &&7352 &&&&&&\\
$4^3F_{4}$ & $4^{+}$ &&7350 &&&&&&\\
$5^3F_{2}$ & $2^{+}$ && 7582 &&&&&&\\
$5^1F_{3}$ & $3^{+}$&& 7579 &&&&&&\\
$5^3F_{3}$ & $3^{+}$&& 7579 &&&&&&\\
$5^3F_{4}$ & $4^{+}$&&7577 &&&&&&\\
\hline
\end{tabular}
}
\end{center}
\end{table*}

\section{Regge Trajectories}
\label{sec:Regge}
\begin{figure}
  \centering
\includegraphics[scale=0.4]{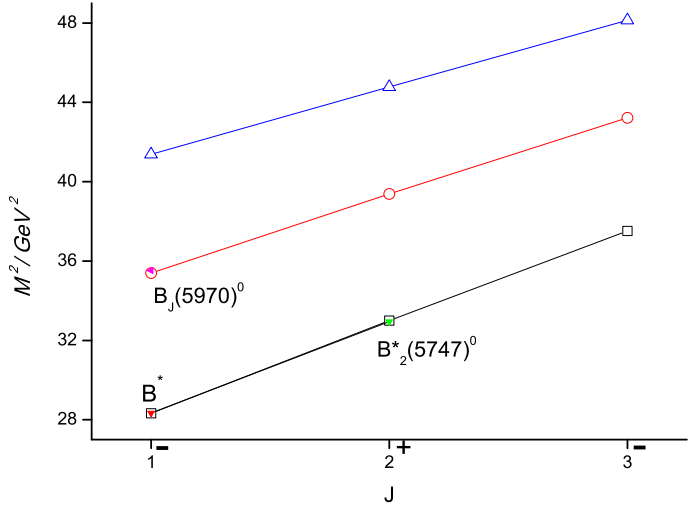}
\caption{\label{fig:natural} $J \rightarrow M^{2}$ Regge trajectory for $B$-meson, natural parity states, hollow shapes indicates present results and solid shapes indicates experimental results}
\end{figure}

\begin{figure}
  \centering
\includegraphics[scale=0.4]{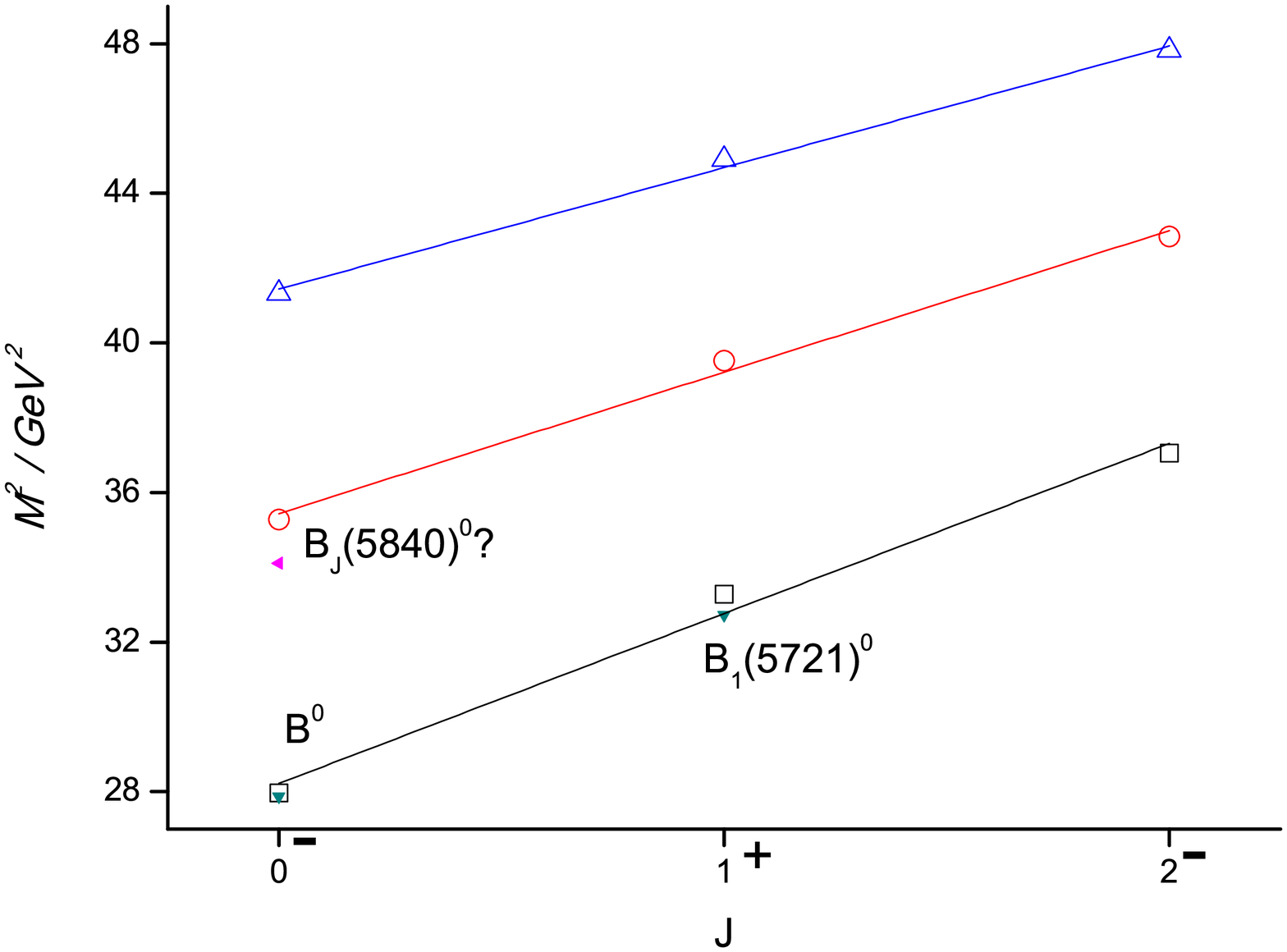}
\caption{\label{fig:un} $J \rightarrow M^{2}$ Regge trajectory for $B$-meson, un-natural parity states, hollow shapes indicates present results and solid shapes indicates experimental results}
\end{figure}

\begin{figure}
  \centering
\includegraphics[scale=0.4]{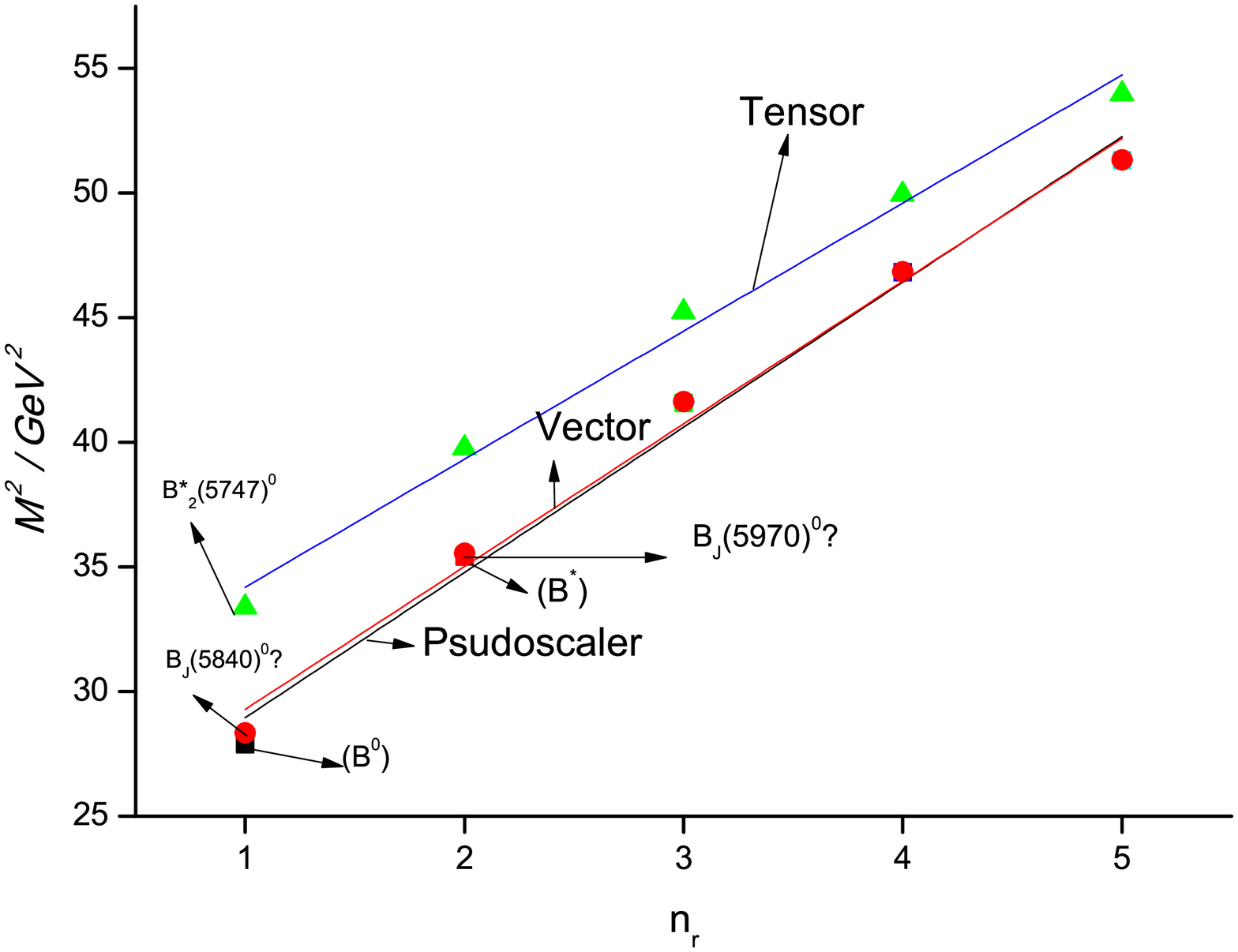}
\caption{\label{fig:ps} $n_r \rightarrow M^{2}$ Regge trajectory for $B$-meson, pseudoscalar, vector and tensor states}
\end{figure}

\begin{figure}
  \centering
\includegraphics[scale=0.4]{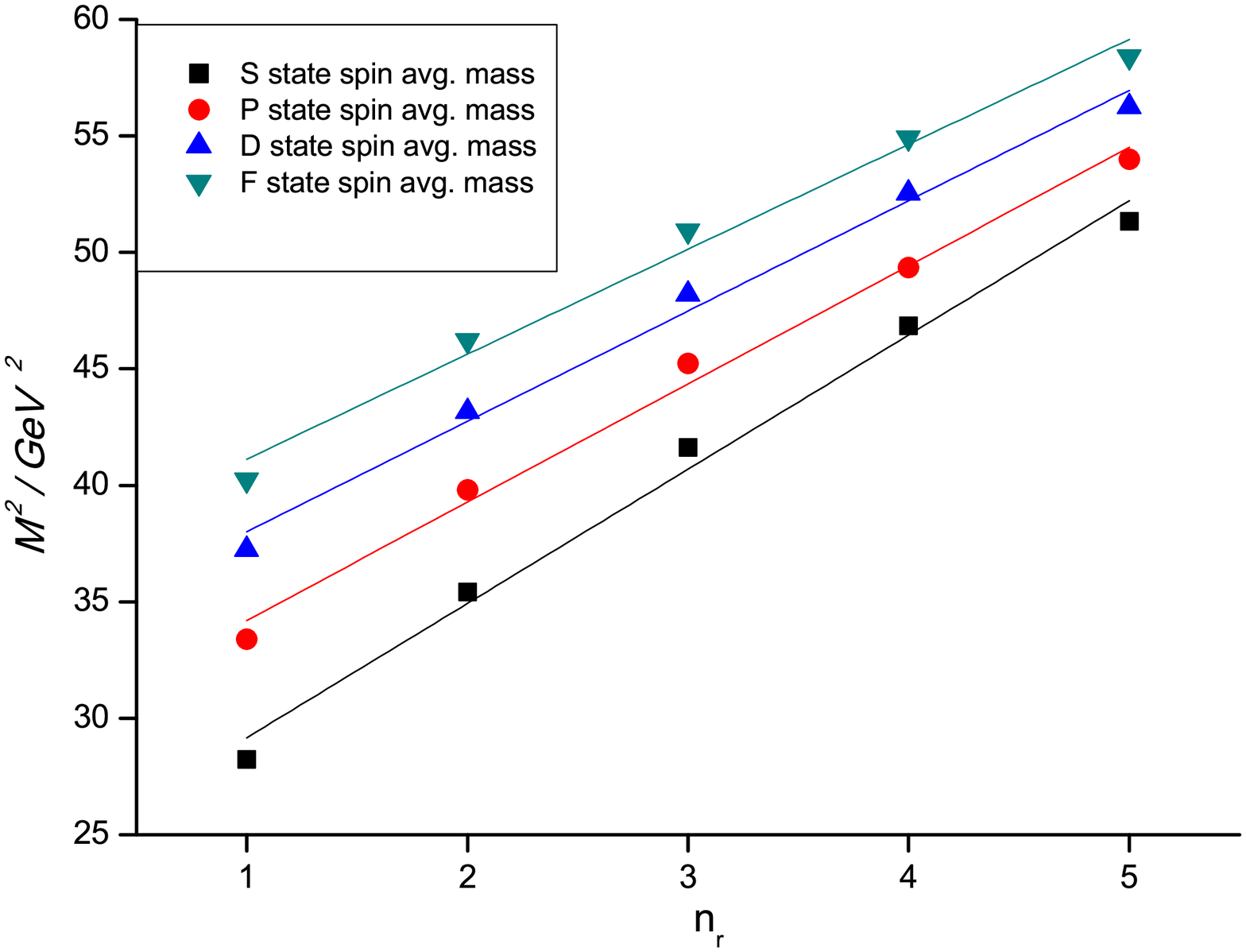}
\caption{\label{fig:sa} $n_r \rightarrow M^{2}$ Regge trajectory for $B$-meson, spin-average masses}
\end{figure}

\begin{figure}
  \centering
\includegraphics[scale=0.4]{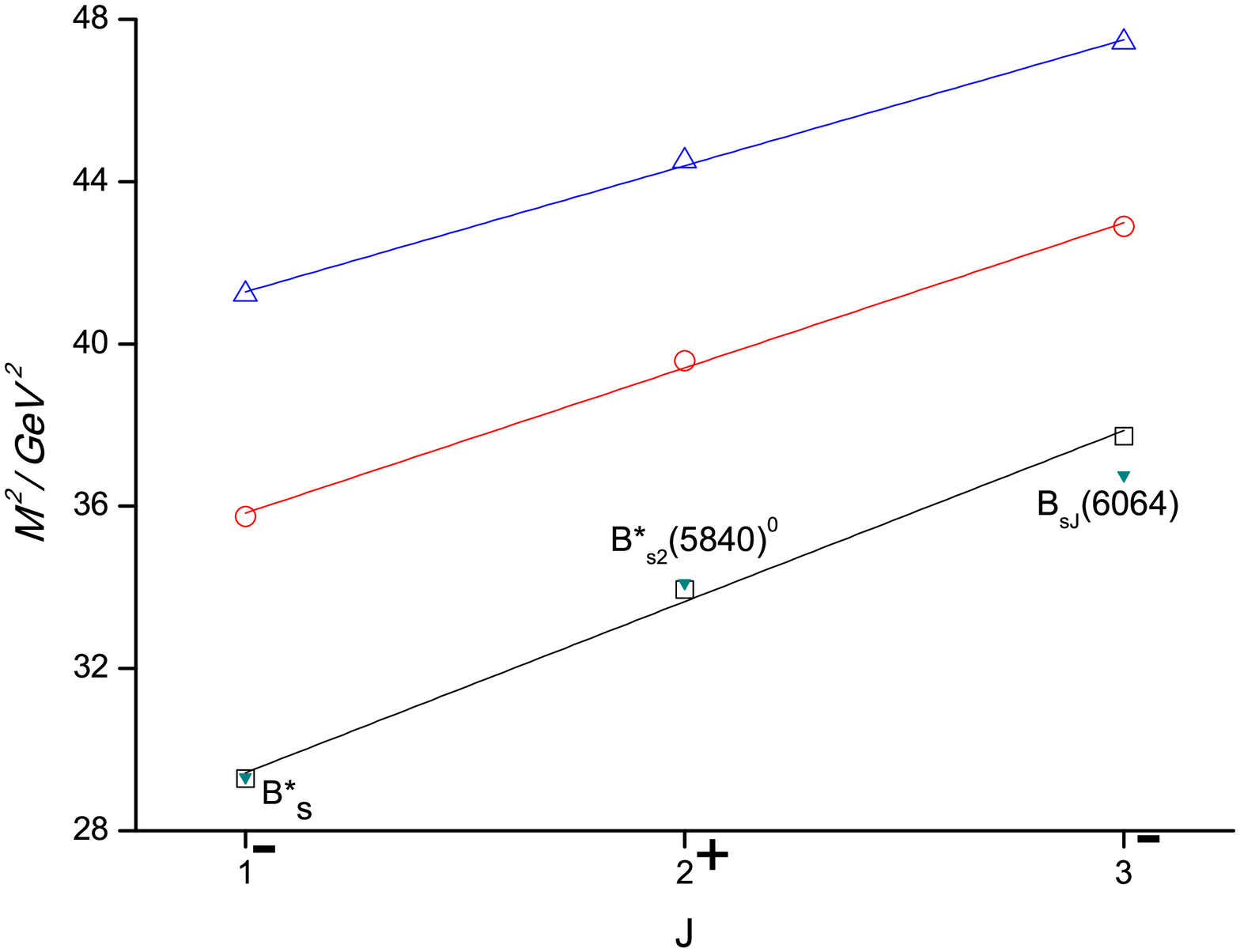}
\caption{\label{fig:natural1} $J \rightarrow M^{2}$ Regge trajectory for $B_s$-meson, natural parity states, hollow shapes indicates present results and solid shapes indicates experimental results}
\end{figure}

\begin{figure}
  \centering
\includegraphics[scale=0.4]{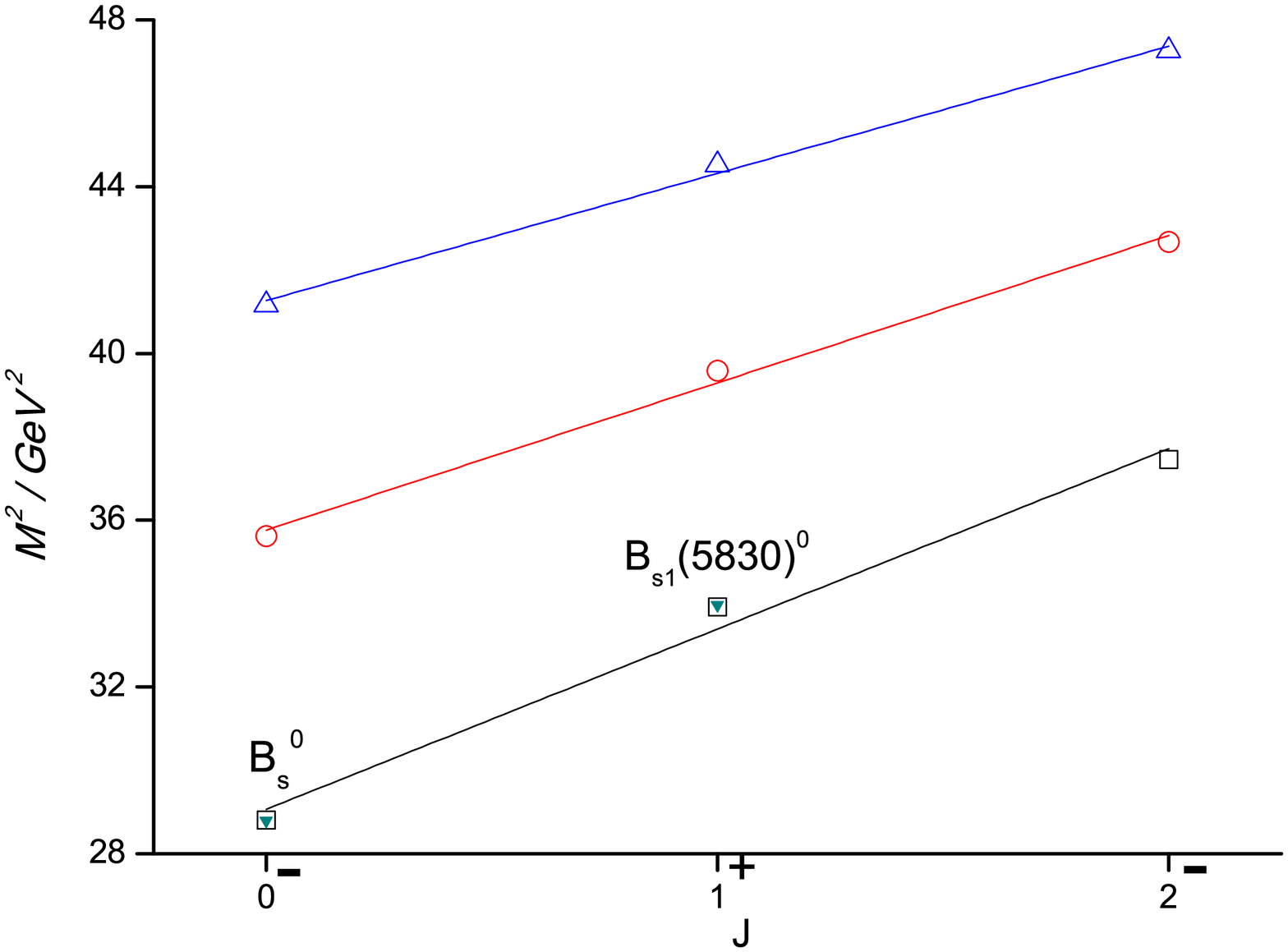}
\caption{\label{fig:un1} $J \rightarrow M^{2}$ Regge trajectory for $B_s$-meson, natural parity states, hollow shapes indicates present results and solid shapes indicates experimental results}
\end{figure}

\begin{figure}
  \centering
\includegraphics[scale=0.4]{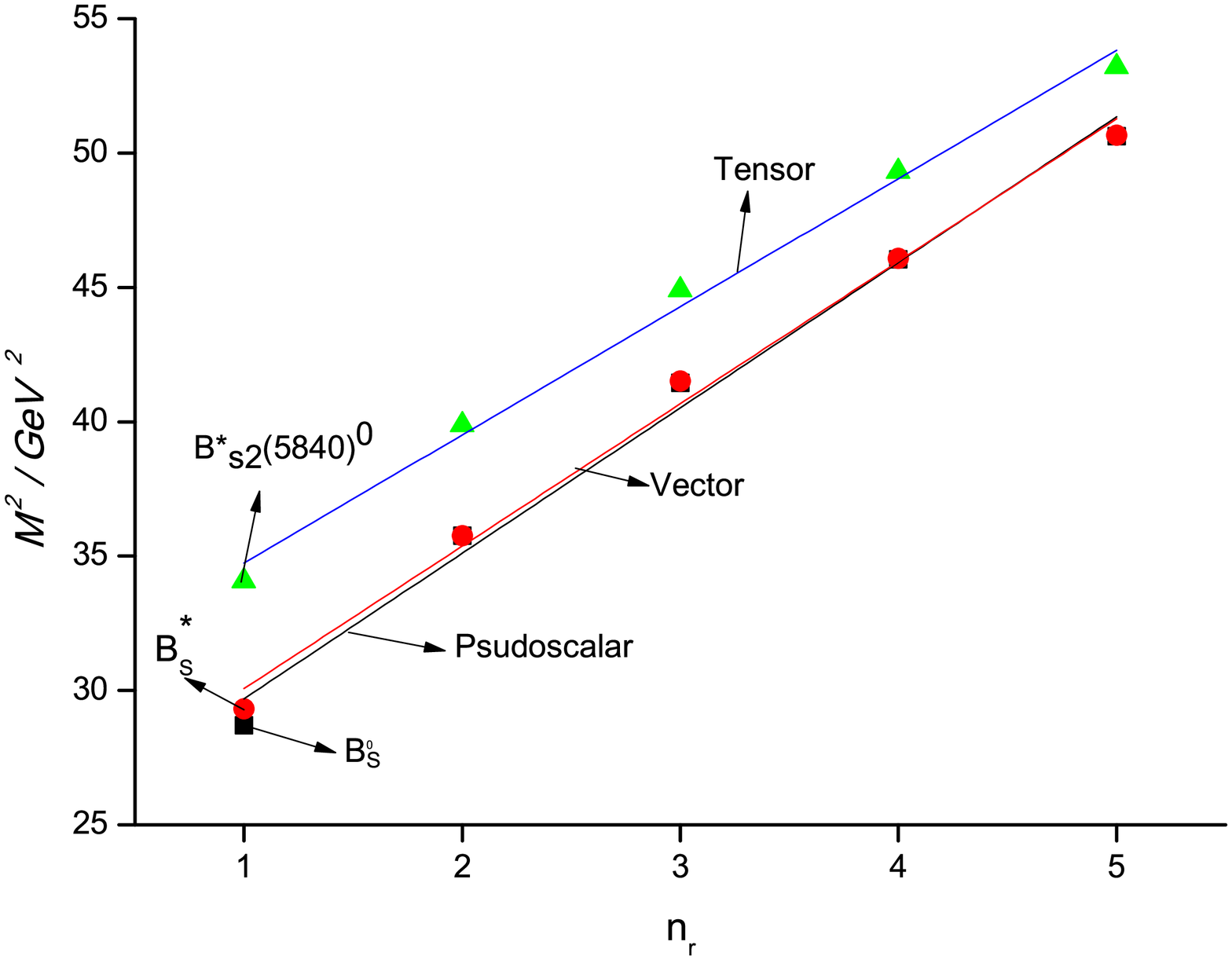}
\caption{\label{fig:ps1} $n_r \rightarrow M^{2}$ Regge trajectory for $B_s$-meson, , pseudoscalar, vector and tensor states}
\end{figure}

\begin{figure}
  \centering
\includegraphics[scale=0.4]{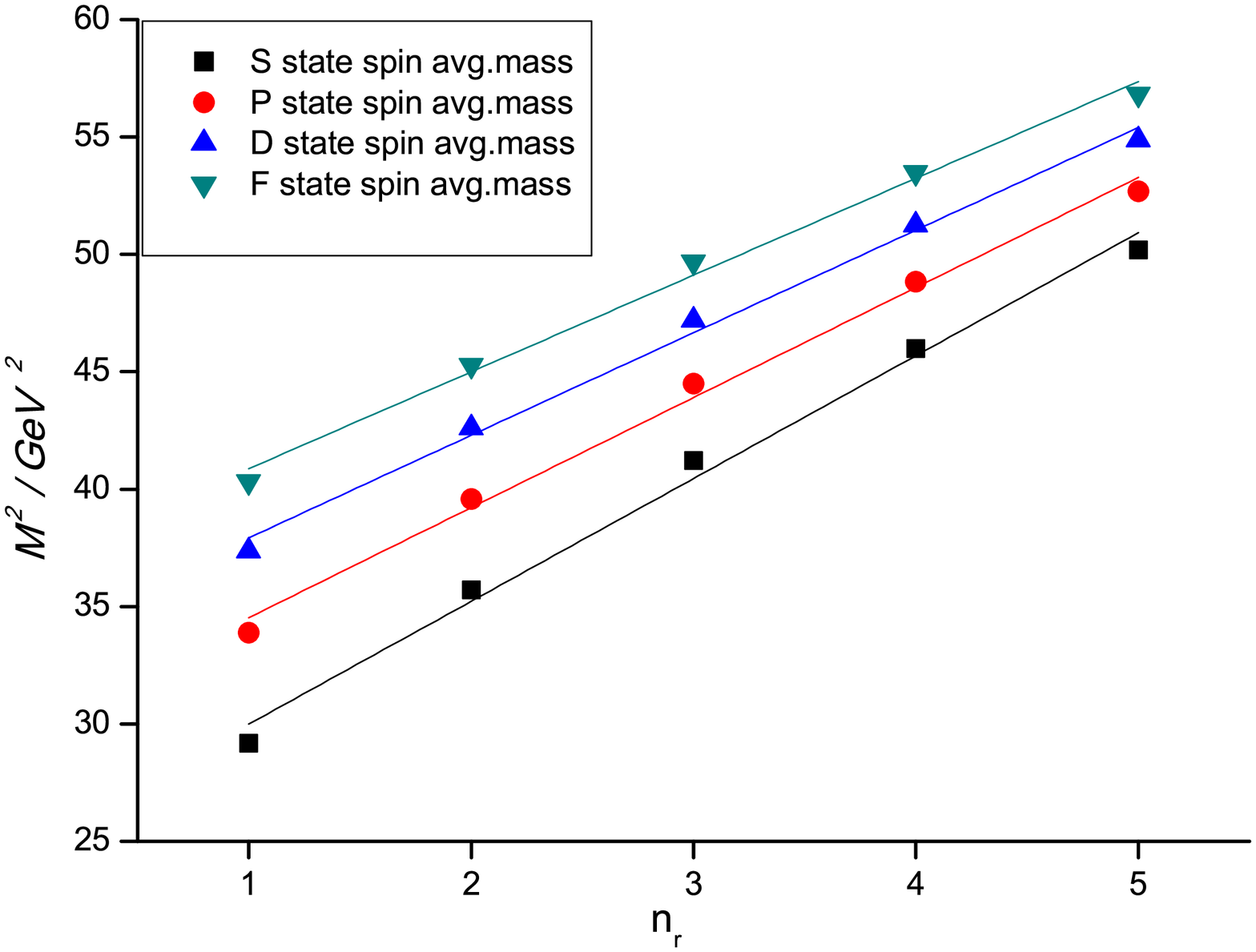}
\caption{\label{fig:sa1} $J \rightarrow M^{2}$ Regge trajectory for $B_s$-meson, , spin-average masses}
\end{figure}

Using the calculated masses from the present approach we build up Regge trajectories in $(J \rightarrow M^{2})$ and $(n_r \rightarrow M^{2})$ planes. To construct the Regge trajectories we use the definitions $J=\alpha M^{2}+\alpha_{0}$ and $n_{r}\equiv n-1=\beta M^{2}+\beta_{0}$. Here, $\alpha_{0},$ $\beta_{0}$ are intercepts, and $\alpha,$ $\beta$ are the slopes. The natural and un-natural parity Regge trajectories in for $(J \rightarrow M^{2})$ planes for $B$ and $B_s$ mesons can be found in figures \ref{fig:natural}, \ref{fig:un}, \ref{fig:natural1}, and \ref{fig:un1}. Whereas, the Regge trajectories in $(n_r \rightarrow M^{2})$ planes for $B$ and $B_s$ mesons can be found in figures \ref{fig:ps}, \ref{fig:sa}, \ref{fig:ps1} and \ref{fig:sa1}. The calculated values of slopes and intercepts can be found in Tables \ref{Table:slope1}, \ref{Table:slope2} and \ref{Table:slope3}.

\begin{table*}
   \caption{\label{Table:slope1} Fitted parameters of the $(J,\: M^{2})$ parent and daughter Regge trajectories for $B$ and $B_{s}$ - mesons with natural and un-natural parity.\label{tab:alfa}}
   \noindent \centering{}%
   \begin{tabular}{ccccc}
   \hline
   \noalign{\smallskip}
   \multirow{2}{*}{Parity} & \multirow{2}{*}{Meson} & \multirow{2}{*}{Trajectory} & \multicolumn{1}{c}{\multirow{2}{*}{$\alpha(GeV^{-2})$}} & \multirow{2}{*}{$\alpha_{0}$}\tabularnewline\noalign{\smallskip}

   \noalign{\smallskip}\hline\noalign{\smallskip}
   \multirow{3}{*}{Natural} & \multirow{3}{*}{$B$} & Parent & $0.217\pm0.002$ & $-5.167\pm0.067$\tabularnewline
    &  & $1^{st}$ daughter & $0.255\pm0.003$ & $-8.048\pm0.125$\tabularnewline
    &  & $2^{nd}$ daughter & $0.296\pm0.001$ & $-11.260\pm0.052$\tabularnewline

   \noalign{\smallskip}

   \hline
   \noalign{\smallskip}
   \multirow{3}{*}{Un-natural } & \multirow{3}{*}{$B^{*}$}& Parent & $0.217\pm0.021$ & $-6.14\pm0.707$\tabularnewline
    &  & $1^{st}$ daughter & $0.263\pm0.018$ & $-9.315\pm0.725$\tabularnewline
    &  & $2^{nd}$ daughter & $0.306\pm0.017$ & $-12.706\pm0.784$\tabularnewline

   \noalign{\smallskip}

    \noalign{\smallskip}\hline\noalign{\smallskip}
   \multirow{3}{*}{Natural} & \multirow{3}{*}{$B_{s}$} & Parent & $0.236\pm0.014$ & $-5.951\pm0.477$\tabularnewline
    &  & $1^{st}$ daughter & $0.279\pm0.011$ & $-8.998\pm0.453$\tabularnewline
    &  & $2^{nd}$ daughter & $0.321\pm0.001$ & $-12.279\pm0.466$\tabularnewline

   \noalign{\smallskip}

   \hline
   \noalign{\smallskip}
   \multirow{3}{*}{Un-natural } & \multirow{3}{*}{$B_{s}^{*}$}& Parent & $0.228\pm0.024$ & $-6.638\pm0.811$\tabularnewline
    &  & $1^{st}$ daughter & $0.281\pm0.020$ & $-10.069\pm0.799$\tabularnewline
    &  & $2^{nd}$ daughter & $0.326\pm0.019$ & $-13.48\pm0.859$\tabularnewline

   \noalign{\smallskip}

   \hline
   \end{tabular}
   \end{table*}

 \begin{table*}
 \caption{\label{Table:slope2} Fitted parameters of the $(n_{r},\: M^{2})$ Regge trajectories for$B$ and $B_{s}$ -mesons \label{tab:bita}}
 \noindent \begin{centering}
 \begin{tabular}{ccccc}
 \hline \noalign{\smallskip}
Mesons & $\beta(GeV^{-2})$ & $\beta_{0}$\tabularnewline
 \noalign{\smallskip}
 \hline
 \noalign{\smallskip}
 %$c\bar{q}$ &  & \tabularnewline
$B^0$ & $0.169\pm0.009$ & $-3.869\pm0.413$\tabularnewline
$B^{*}$& $0.173\pm0.009$ & $-4.050\pm0.388$\tabularnewline
$B^{*}_{2}(5747)^0$ & $0.193\pm0.009$ & $-5.58\pm0.448$\tabularnewline
 \hline
 $B_{s}^0$ & $0.183\pm0.010$ & $-4.416\pm0.423$\tabularnewline
$B_{s}^{*}$& $0.187\pm0.008$ & $-4.623\pm0.360$\tabularnewline
$B^{*}_{s2}(5840)^0$ & $0.208\pm0.009$ & $-6.627\pm0.424$\tabularnewline
\hline
 \end{tabular}
 \par\end{centering}
 \end{table*}

\begin{table*}
 \caption{\label{Table:slope3} Fitted parameters of the $(n_{r},\: M^{2})$ S, P, D and F state spin-average mass Regge trajectories for $B$ and $B_{s}$ - mesons.\label{tab:Spinave} }

 \noindent \centering{}%
 \begin{tabular}{ccccc}
 \hline\noalign{\smallskip}

Meson & Trajectory & $\beta(GeV^{-2})$ & $\beta_{0}$\tabularnewline

 \noalign{\smallskip}\hline\noalign{\smallskip}

 \multirow{3}{*}{$B$} & S State & $0.155\pm0.015$ & $-4.187\pm0.642$\tabularnewline
  & P State & $0.195\pm0.01$ & $-6.631\pm0.446$ \tabularnewline
  & D State & $0.212\pm0.011$ & $-7.973\pm0.526$\tabularnewline
  & F State & $0.226\pm0.012$ & $-9.273\pm0.638$\tabularnewline
  \hline
  \multirow{3}{*}{$B_{s}$} & S State & $0.189\pm0.009$ & $-5.673\pm0.395$\tabularnewline
  & P State & $0.212\pm0.009$ & $-7.310\pm0.418$ \tabularnewline
  & D State & $0.227\pm0.009$ & $-8.623\pm0.458$\tabularnewline
  & F State & $0.241\pm0.010$ & $-9.841\pm0.538$\tabularnewline

 \noalign{\smallskip}
 \hline
 \medskip
 \end{tabular}
 \end{table*}

\section{Decay Constant}
\label{sec:DecayConstant}
The decay constants are important parameters for studying the leptonic and non-leptonic decay processes. After incorporating the fist order QCD corrections the decay constants for various states have been calculated by using the Van-Royen-Weisskopf formula\cite{VanRoyen:1967nq}.
\begin{equation}
f_{p/v}^{2}=\frac{12\left|\psi_{P/V}(0)\right|^{2}}{M_{P/V}}\bar{C^{2}}(\alpha_{S}),\label{Eq:decayconst}
\end{equation}
where, \ensuremath{\bar{C^{2}}(\alpha_{S})} is the QCD correction factor given by \cite{Braaten:1995ej}. The calculated values of the decay constant for $B$ and $B_s$ meson can be found in Table \ref{tab:decayconstant}.
\begin{table*}
\begin{center}{\caption{\label{tab:decayconstant} {Decay Constant of pseudoscalar and vector states of $B$ and $B_{s}$ meson (in GeV).}}}
\vspace{0.1cm}
\begin{tabular}{c c c c c c c c c c }
\hline
  &Model & 1S & 2S & 3S & 4S & 5S  \\
\hline
 &f$ _{Pcor}$& 0.149 &0.071  &0.045 &0.033  &0.025  \\
 &f$ _{p} $ & 0.154   &0.073  &0.046  &0.034  &0.026   & \\
 &\cite{tanabashi2018particle} &0.188$\pm$ 25 \\
%    &\cite{patrignani2016review} &0.198$\pm$ 0.014 \\
$B$&\cite{Kher:2017mky} & 0.146 &0.105& 0.086&0.074&0.066& \\
%&\cite{devlani2011spectroscopy}&0.216 &0.101&0.072&0.058 \\
%&\cite{hwang1997decay} & 0.231$\pm$ 0.009 \\
%&\cite{shah2016spectroscopy}&0.188\\
&\cite{capstick1990pseudoscalar}&0.155$\pm$0.015\\
%& \cite{olive2014review}& 0.190$\pm$0.04 & \\
%& \cite{lucha2011ope,lucha2011decay}& 0.193$\pm$0.012 & \\
%& \cite{wang2015analysis}& 0.192$\pm$0.013 & \\
%& \cite{dowdall2013b,colquhoun2015b}& 0.186 & \\
\hline
&f$ _{Vcor} $ & 0.150 &0.071  &0.045  &0.033  &0.025 \\
& f$ _{V} $ & 0.155  &0.071  &0.046  & 0.034 &0.026  \\
$B^{*}$&\cite{Kher:2017mky}& 0.146 &0.105& 0.086&0.074&0.066&  \\
% &\cite{devlani2011spectroscopy}&0.218 &0.101&0.072&0.058 \\
%&\cite{hwang1997decay} & 0.252$\pm$ 0.010 \\
%& \cite{lucha2014decay,lucha2015accurate} & 0.181$\pm$ 0.013 \\
% & \cite{wang2015analysis}&0. 213$\pm$0.018 & \\
% & \cite{ebert2006relativistic}& 0.219 & \\
% & \cite{badalian2007decay}& 0.200$\pm$0.010& \\
 & \cite{verma2012decay}& 0.196 & \\
 & \cite{dowdall2013b,colquhoun2015b}& 0.175 & \\
\hline
 &f$ _{Pcor} $ &0.191  &0.089 &0.0570  &0.0413  &0.0327  \\
 &f$ _{p} $ &0.210   &0.098  &0.0628  &0.0454  &0.0359   \\
 &\cite{patrignani2016review} &0.237$\pm$ 0.0147 \\
$B_{s}$&\cite{Kher:2017mky} &0.187&0.134&0.110&0.095&0.085 \\
 &\cite{PhysRevD.84.074030}&0.232 &0.107&0.077&0.062 \\
%&\cite{hwang1997decay} & 0.266$\pm$ 0.010 \\
%&\cite{shah2016spectroscopy}&0.240\\
&\cite{capstick1990pseudoscalar}&0.210$\pm$0.020\\
%& \cite{lucha2011ope,lucha2011decay}& 0.232$\pm$0.018 & \\
%& \cite{wang2015analysis}& 0.230$\pm$0.013 & \\
%& \cite{dowdall2013b,colquhoun2015b}& 0.224 & \\
\hline
&f$ _{Vcor} $ &0.192  &0.089 &0.0572  &0.0413  &0.0327   \\
& f$ _{V} $ &0.211   &0.098  &0.0629  &0.0454  &0.0359  \\
$B^{*}_{s}$&\cite{Kher:2017mky}&0.187&0.134&0.110&0.095&0.085\\
% &\cite{devlani2011spectroscopy}&0.234 &0.107&0.077&0.062 \\
%  &\cite{hwang1997decay} & 0.289$\pm$ 0.011 \\
  & \cite{lucha2014decay,lucha2015accurate} & 0.213$\pm$ 0.018 \\
%  & \cite{wang2015analysis}& 0.255$\pm$19 & \\
%   & \cite{ebert2006relativistic}&0. 251 & \\
% & \cite{badalian2007decay}&0. 230$\pm$0.012& \\
% & \cite{verma2012decay}& 0.229 & \\
 & \cite{dowdall2013b,colquhoun2015b}& 0.213& \\
   \hline
\end{tabular}
\end{center}
\end{table*}

\section{Leptonic, Radiative leptonic and Dileptonic Branching fractions}
\label{sec:leptonic branching fractions}
The leptonic branching fraction is calculated using the formula,
\begin{equation}
BR=\Gamma\times\tau
\end{equation}
The leptonic decay width($\Gamma$) is given as per\cite{Silverman:1988gc}. Calculated leptonic branching fraction can be found in Table \ref{tab:tableBF}.
\begin{eqnarray}
\Gamma({B}^{+}\rightarrow l^{+}\nu_{l})& = & \frac{G_{F}^{2}}{8\pi}f_{B}^{2}\left|V_{ub}\right|^{2}m_{l}^{2}\times \left(1-\frac{m_{l}^{2}}{M_{B}^{2}}\right)^{2}M_{B}\label{Eq:branchingB}
\end{eqnarray}
The radiative leptonic decay width $B^{-}\rightarrow\gamma l\bar{\nu}\,(l=e,\mu)$ is calculated as per\cite{Lu:2002mn}. The results can be found in Table \ref{tab:tablerld}.
\begin{equation}
\Gamma(B^{-}\rightarrow\gamma l\bar{\nu})=\frac{\alpha G_{F}^{2}\left|V_{bu}\right|^{2}}{2592\pi^{2}}f_{B^{-}}^{2}m_{B^{-}}^{3}\left[x_{u}+x_{b}\right],\label{Eq:radiative}
\end{equation}
where
\begin{equation}
x_{u}=\left(3-\frac{m_{B^{-}}}{m_{u}}\right)^{2},
\end{equation}
and
\begin{equation}
 x_{b}=\left(3-2\frac{m_{B^{-}}}{m_{b}}\right)^{2}
\end{equation}
The rare di-leptonic decay width for $B_{s}^{0}$ and $B^{0}$ mesons is given by\cite{Bobeth:2013uxa,Bobeth:2013tba}. Our calculated rare dileptonic decay width can be found in Tables \ref{tab:rare1} \& \ref{rare2}.
\begin{eqnarray}
\Gamma(B_{q}^{0}\rightarrow l^{+}l^{-})=\frac{G_{F}^{2}}{\pi}\left(\frac{\alpha}{4\pi sin^{2}\Theta_{W}}\right)^{2}f_{B_{q}}^{2}m_{l}^{2}m_{B_{q}}
\times\sqrt{1-4\frac{m_{l}^{2}}{m_{B_{q}}^{2}}}|V_{tq}^{\star}V_{tb}|^{2}|C_{10}|^{2}\label{Eq:rareB}
\end{eqnarray}
The branching ratio for $B_{q}^{0}\rightarrow l^{+}l^{-}$ is
\begin{equation}
BR\rightarrow\Gamma_{(B_{q}^{0}\rightarrow l^{+}l^{-})}\times\tau_{B_{q}}
\end{equation}
$C_{10}$ is the Wilson coefficient given by \cite{Shah:2016mgq,Buchalla:1993bv}, $f_{B_{q}}$ is the corresponding decay constant, and $G_{F}$ is the Fermi coupling constant.  We have considered $\tau_{B}=1.638$ ps and  $\tau_{B_s}=1.510$ ps.
\begin{equation}
C_{10}=\eta_{Y}\frac{x_{t}}{8}\left[\frac{x_{t}+2}{x_{t}-1}+\frac{3x_{t}-6}{(x_{t}-1)^{2}}ln\: x_{t}\right]
\end{equation}
$\Theta_{W}(\approx28^{0})$ is the weak mixing angle (Weinberg angle)\cite{Lee:2015yht}, $\eta_{Y}(=1.026)$ is the next-to-leading-order correction\cite{Shah:2016mgq}, and $x_{t}=\left(m_{t}/m_{w}\right)^{2}$.

\begin{table}[!h]
\begin{center}
{\caption{\label{tab:tableBF} {Leptonic Branching fraction of the $B$ meson.}}}
\vspace{0.1cm}
\begin{tabular}{c c c c}
\hline
  State &$ B^{+}\rightarrow\tau^{+}\nu_{\tau}$ & $ B^{+}\rightarrow\mu^{+}\nu_{\mu}$ & $ B^{+}\rightarrow e^{+}\nu_{e}$   \\
 & $Br_{\tau}$ & $Br_{\mu}$ &$Br_{e}$ \\
\hline
Present&0.0285$\times 10^{-4}$ &0.128$\times 10^{-7}$&0.3$\times 10^{-12}$   \\
\cite{tanabashi2018particle}&(1.09$\pm$0.24)$\times 10^{-4}$  &$<$1.0$\times 10^{-6}$  &$<$9.8$\times 10^{-7}$ \\
 & &(CL= 90$\%$) &  (CL=90$\%$)\\
\cite{Kher:2017mky}&0.822$\times 10^{-4}$ & 0.37$\times 10^{-7}$ &8.64$\times 10^{-12}$ \\
\cite{shah2016spectroscopy} &1.354$\times 10^{-4}$  \\
\cite{fu2012study} &1.1$\times 10^{-4}$   &4.7$\times 10^{-7}$ & 1.11$\times 10^{-11}$\\
\cite{PhysRevD.84.074030} &1.13$\times 10^{-11}$   &4.82$\times 10^{-7}$ & 1.07$\times 10^{-7}$\\

\hline
\end{tabular}
\end{center}
\end{table}

%\begin{table}[!h]
%\begin{center}
%{\caption{\label{tab:tablerld} {Radiative leptonic decay width and branching ratio of $B$ and $B_{s}$ }}}
%\vspace{0.1cm}
%\begin{tabular}{c c c c c c c c c c c}
%\hline
 % Meson & Decay Constant & $ \Gamma/GeV$ &&& $BR$\\
  %\hline
%& & &Present & \cite{Kher:2017mky}&  \cite{korchemsky2000radiative}&\cite{yang2014factorization}&\cite{yang2017factorization}\\
%\hline
%$B$&$fp$ &1.23$ \times 10^{-19} $&0.30$ \times 10^{-6} $&0.38$ \times 10^{-6} $&0.23$ \times 10^{-6} $&1.66$ \times 10^{-6} $&5.21$ \times 10^{-6} $\\
%&$fpcor$ &0.19$ \times 10^{-19} $&0.47$ \times 10^{-7} $&0.34$ \times 10^{-6} $&\\
%\hline
 % $B_{s}$&$fp$ &2.49$ \times 10^{-19} $&6.21$ \times 10^{-7} $&\\
  %&$fpcor$ &5.24$ \times 10^{-20} $&1.30$ \times 10^{-7} $&\\
%\hline
%\end{tabular}
%\end{center}
%\end{table}

\begin{table}[!h]
\begin{center}
{\caption{\label{tab:tablerld} {Branching ratio with corresponding radiative leptonic decay width }}}
\vspace{0.1cm}
\begin{tabular}{ c c c c c c c c c c}
\hline
  Meson & & $ \Gamma/GeV$ &&& $BR$\\
  \hline
&&  &Present & \cite{Kher:2017mky}&  \cite{korchemsky2000radiative}&\cite{yang2014factorization}&\cite{yang2017factorization}\\
\hline
$B$ &$f_{p}$&1.23$ \times 10^{-19} $&0.30$ \times 10^{-6} $&0.38$ \times 10^{-6} $&0.23$ \times 10^{-6} $&1.66$ \times 10^{-6} $&5.21$ \times 10^{-6} $\\
&$f_{pcor}$ &0.19$ \times 10^{-19} $&0.47$ \times 10^{-7} $&0.34$ \times 10^{-6} $&\\
\hline
  $B_{s}$ &$f_{p}$&2.49$ \times 10^{-19} $&6.21$ \times 10^{-7} $&\\
 &$f_{pcor}$ &5.24$ \times 10^{-20} $&1.30$ \times 10^{-7} $&\\
\hline
\end{tabular}
\end{center}
\end{table}

\begin{table}[!h]
\begin{center}
{\caption{\label{tab:rare1} {Branching ratio with corresponding rare leptonic decay width of the $B^{0}$ meson.}}}
\vspace{0.1cm}
\begin{tabular}{c c c c c c c c c}
\hline
Process&$\Gamma(B^{0}_{q}\rightarrow l^{+} l^{-})(keV)$&& $BR$&\\
&Present&Others&Present&Others&\\
\hline
 $ B^{0}\rightarrow\mu^{+}\mu^{-}$ &2.49$\times 10^{-17}$ & 4.34$\times 10^{-17}$  \cite{Kher:2017mky}&5.75$\times 10^{-11}$&$(3.9^{+1.6}_{-1.4})\times 10^{-10}$\cite{patrignani2016review}\\
&&4.41$\times 10^{-17}$\cite{shah2016spectroscopy}&&1.018$\times 10^{-10}$\cite{shah2016spectroscopy}\\
&&&&1.00$\times 10^{-10}$ \cite{Kher:2017mky}\\
&&&&$<$1.1$\times 10^{-9}$ \cite{chatrchyan2013measurement}\\
%&&&&$<$9.4$\times 10^{-10}$ \cite{aaij2013first}\\
%&&&&$<$7.4$\times 10^{-10}$ \cite{aaij2013measurement}\\
%&&&&1.20$\times 10^{-10}$ \cite{dimopoulos2012lattice}\\
%&&&&$(1.06\pm 0.09)\times 10^{-10}$ \cite{bobeth2014b}\\
\hline
$ B^{0}\rightarrow\tau^{+}\tau^{-}$& 5.22$\times 10^{-15}$& 9.10$\times 10^{-10}$\cite{Kher:2017mky}&1.20$\times 10^{-8}$&$<$4.1$\times 10^{-3}$\cite{patrignani2016review}\\
&&9.23$\times 10^{-10}$\cite{shah2016spectroscopy}& &2.133$\times 10^{-8}$\cite{shah2016spectroscopy}\\
&&& &2.10$\times 10^{-8}$\cite{Kher:2017mky}\\
&&& &$(2.22\pm 0.19)\times 10^{-8}$ \cite{bobeth2014b}\\
&&&&2.52$\times 10^{-8}$ \cite{dimopoulos2012lattice}\\

\hline

  $ B^{0}\rightarrow e^{+}e^{-}$ & 5.84$\times 10^{-22}$& 1.02$\times 10^{-21}$\cite{Kher:2017mky}&1.34$\times 10^{-15}$&$<$8.3$\times 10^{-8}$\cite{patrignani2016review}\\
  &&1.03$\times 10^{-21}$\cite{shah2016spectroscopy}& &2.376$\times 10^{-15}$\cite{shah2016spectroscopy}\\
  &&& &2.35$\times 10^{-15}$\cite{Kher:2017mky}\\
&&& &$(2.48\pm 0.21)\times 10^{-15}$ \cite{bobeth2014b}\\
&&&&2.82$\times 10^{-15}$ \cite{dimopoulos2012lattice}\\

\hline
\end{tabular}
\end{center}
\end{table}

\begin{table}[!h]
\begin{center}
{\caption{\label{rare2} {Branching ratio with corresponding rare leptonic decay width of the $B^{0}_{s}$ meson.}}}
\vspace{0.1cm}
\begin{tabular}{c c c c c c c c c}
\hline
Process&$\Gamma(B^{0}_{q}\rightarrow l^{+} l^{-})(keV)$&& $BR$&\\
&Present&Others&Present&Others&\\
\hline
 $B^{0}_{s}\rightarrow\mu^{+}\mu^{-}$ &4.70$\times 10^{-17}$ & 1.10$\times 10^{-15}$  \cite{Kher:2017mky}&1.08$\times 10^{-10}$&$(2.9^{+0.7}_{-0.6})\times 10^{-9}$\cite{patrignani2016review}\\
&&1.58$\times 10^{-15}$\cite{shah2016spectroscopy}&&3.602$\times 10^{-9}$\cite{shah2016spectroscopy}\\
&&&&2.53$\times 10^{-9}$ \cite{Kher:2017mky}\\
&&&&$(3.0^{+1.0}_{-0.9})\times 10^{-9}$ \cite{chatrchyan2013measurement}\\
%&&&&$(3.2^{+1.5}_{-1.2})\times 10^{-9}$\cite{aaij2013first}\\
&&&&$(2.9^{+1.1}_{-1.0})\times 10^{-9}$\cite{aaij2013measurement}\\
%&&&&3.40$\times 10^{-9}$ \cite{dimopoulos2012lattice}\\
%&&&&$(3.65\pm 0.23)\times 10^{-9}$ \cite{bobeth2014b}\\
\hline
$ B^{0}_{s}\rightarrow\tau^{+}\tau^{-}$& 9.97$\times 10^{-15}$& 2.34$\times 10^{-13}$\cite{Kher:2017mky}&2.30$\times 10^{-8}$&7.647$\times 10^{-7}$ \cite{dimopoulos2012lattice}\\
&&3.36$\times 10^{-13}$\cite{shah2016spectroscopy}& &7.647$\times 10^{-7}$\cite{shah2016spectroscopy}\\
&&& &5.36$\times 10^{-7}$\cite{Kher:2017mky}\\
&&& &$(7.73\pm 0.0.23)\times 10^{-7}$ \cite{bobeth2014b}\\

\hline

  $ B^{0}_{s}\rightarrow e^{+}e^{-}$ & 1.10$\times 10^{-21}$& 2.58$\times 10^{-21}$\cite{Kher:2017mky}&2.54$\times 10^{-15}$&$<$2.8$\times 10^{-7}$\cite{patrignani2016review}\\
  &&3.70$\times 10^{-20}$\cite{shah2016spectroscopy}& &8.408$\times 10^{-14}$\cite{shah2016spectroscopy}\\
  &&& &5.92$\times 10^{-14}$\cite{Kher:2017mky}\\
&&& &$(8.54\pm 0.55)\times 10^{-14}$ \cite{bobeth2014b}\\
&&&&7.97$\times 10^{-14}$ \cite{dimopoulos2012lattice}\\

\hline
\end{tabular}
\end{center}
\end{table}

\section{Mixing Parameters}
\label{sec:mix}
Many experiments\cite{Abulencia:2006mq,Komiske:2020qhg} have reported neutral open beauty meson oscillations. Using the spectroscopic parameters from the present study we calculate the mass oscillation of the neutral open beauty meson and integrated oscillation rate. We use the notation available in\cite{Zyla:2020zbs} and also consider CPT conservation. Time evolution of the neutral $B$ and $B_s$ meson doublet is described by the Schrodinger equation\cite{Buchalla:2008jp}.
\begin{equation}
i\frac{d}{dt}\left(\frac{D_{q}}{D_{q}}\right)=\left[\left(\begin{array}{cc}
M_{11}^{q} & M_{12}^{q\star}\\
M_{12}^{q} & M_{11}^{q}
\end{array}\right)-\frac{i}{2}\left(\begin{array}{cc}
\Gamma_{11}^{q} & \Gamma_{12}^{q\star}\\
\Gamma_{12}^{q} & \Gamma_{11}^{q}
\end{array}\right)\right]\left(\frac{D_{q}}{D_{q}}\right) \label{Eq:bbr}
\end{equation}
The off-diagonal elements of the mass and decay matrices are\cite{Buras:1984pq}
\begin{equation}
M_{12}=-\frac{G_{F}^{2}m_{W}^{2}\eta_{D}m_{D_{q}}B_{D_{q}}f_{D_{q}}^{2}}{12\pi^{2}}S_{0}\left(m_{s}^{2}/m_{W}^{2}\right)\left(V_{us}^{\star}V_{cs}\right)^{2}
\end{equation}
\begin{equation}
\Gamma_{12}=\frac{G_{F}^{2}m_{c}^{2}\eta_{D}^{\prime}m_{D_{q}}B_{D_{q}}f_{D_{q}}^{2}}{8\pi}\left[\left(V_{us}^{\star}V_{cs}\right)^{2}\right]
\end{equation}
Here, $m_W$ is the W boson mass, $m_b$ is the mass of $b$-quark, $G_f$ is Fermi constant, and $m_{B_{q}}$, $f_{B_{q}}$ and $B_{B_{q}}$ are the ${B_{q}}$ mass, the weak decay constant and the bag parameter, respectively. $V_{ij}$ are the CKM matrix elements and $S_{0}(x_{t})$\cite{Kobayashi:1973fv} can be approximated as $0.784x_{t}^{0.76}$\cite{Cichy:2016bci}. $\eta_{D}$ and $\eta_{D}^{\prime}$ resemble gluonic corrections. $\tau_{B}$ (hadronic lifetime) is related to $\Gamma_{11}^{q}\left(\tau_{B_{q}}=1/\Gamma_{11}^{q}\right)$, and $\Delta m_{q}$ and $\Delta\Gamma_{q}$ are related to $M_{12}^{q}$ and $\Gamma_{12}^{q}$\cite{Tanabashi:2018oca}.
\begin{equation}
\triangle m_{q}=2\left|M_{12}^{q}\right|
\end{equation}
and
\begin{equation}
\triangle\Gamma_{q}=2\left|\Gamma_{12}^{q}\right|
\end{equation}
$\chi_{q}$(integrated oscillation rate) is the probability of observing a $\bar{B_{q}}$ meson in a jet initiated by $\bar{b}$ quark, $\Delta m_{B_q}$ is a measure of frequency of the change from a $B_q$ into a $\bar{B_q}$ or vice versa.\\
The ratio;
\begin{equation} \label{cosh5}%arXiV% $$
 r_o = \frac{B_q \leftrightarrow \bar{B_q}}{B_q \leftrightarrow B_q} = \frac{\int^\infty_0 \left|g_- (t) \right|^2 dt}{\int^\infty_0 \left|g_+ (t) \right|^2 dt} = \frac{x^2}{2 + x^2},
\end{equation}
\begin{equation}
 \left|g_\pm (t) \right|^2\ = \frac{1}{2} e^{ \frac{-\Gamma_D t}{2}}  [1 \pm   \cos(t \Delta
m)].
\end{equation}
\begin{eqnarray} \label{chi} %arXiV% $$
~~~{\rm where}~~~ x_q = x =
\frac{\Delta m}{\Gamma}= \Delta m\ \tau_{D},  ~~~ y_q =
\frac{\Delta \Gamma }{2\Gamma}=\frac{\Delta \Gamma
\ \tau_{D}}{2}    \nonumber,
\end{eqnarray} %arXiV% $$
\begin{equation}
 \chi_q = \frac{x_q^2+y_q^2}{2(x_q^2+1)},
\end{equation}
In the absence of CP violation, we have the time-integrated mixing rate for semi-leptonic decays as
\begin{equation}
R_{M}\simeq\frac{1}{2}(x_{q}^{2}+y_{q}^{2}).
\end{equation}
For present computation of the mixing parameters, we employ our calculated value for $\Delta m$ and experimental average lifetime from PDG. The computed values of mixing parameters can be found in Table \ref{tab:mix}.
\begin{table}
\begin{center}
{\caption{\label{tab:mix}{Mixing of $B$ and $B_{s}$ mesons variables.}}}
\vspace{0.1cm}
\resizebox{\textwidth}{!}{
\begin{tabular}{ccccccc}
\hline
 States & $\bigtriangleup m_{q}\times10^{-12}$&  $x_{q}$ &$y_{q}$&$\chi_{q}$ & $R_{M}$ \\
\hline
Present  ($B$) &0.25374&0.415627&1.281$\times 10^{-3}$&0.073&0.086&\\
\cite{tanabashi2018particle}&0.5064$\pm$0.0019&0.770$\pm$0.004&&1.860$\pm$0.0011&&\\
\cite{Kher:2017mky}&0.5397&0.8841&2.63$\times 10^{-3}$&0.2199&0.3909\\
\cite{shah2016spectroscopy}&0.506&0.769&&&\\
\cite{PhysRevD.84.074030} &0.498&0.759&2.6$\times 10^{-3}$&0.2242&&\\
%\cite{patel2009mixing}&0.593&0.9014&4.6$\times 10^{-3}$&0.2242&&\\
% \cite{lenz2011numerical}& 0.543$\pm$ 0.091 &&&&\\
\hline
Present  ($B_{s}$)&25.37 &12.68&3.81$\times 10^{-3}$&0.00079 &0.0080 &\\
\cite{tanabashi2018particle}&17.757$\pm$0.021&26.79$\pm$0.08&&0.499307$\pm$0.000004&&\\
\cite{Kher:2017mky}&14.13&21.36&5.40$\times 10^{-2}$&0.4989&228.23\\
\cite{shah2016spectroscopy}&&17.644&26.41&8.9$\times 10^{-2}$&0.4993&\\
%\cite{devlani2011spectroscopy} &11.896&17.511&6.10$\times 10^{-2}$&0.4984&&\\
%\cite{patel2009mixing}&23.36&33.95&17.22$\times 10^{-2}$&0.4996&&\\
\cite{lenz2011numerical}& 17.30$\pm$ 2.6 &&&&\\
 \cite{lahkar2019masses}& 15.8 &&&&\\
%  \cite{lahkar2019masses}& 9.3 &&&&\\
\hline
\end{tabular}
}
\end{center}
\end{table}

\section{Electromagnetic transition widths}
\label{sec:EM}
The electromagnetic transition width help to understand the non-perturbative characteristic of QCD. $\Delta L = \pm 1$ and $\Delta S = 0$ and, $\Delta L = 0$ and $\Delta S = \pm 1$ are the selection rules for calculating the electric and magnetic transition widths. We use the variational radial wave-functions for initial and final states to calculate electromagnetic transition widths. In non-relativistic limit the, widths can be calculated as per\cite{Brambilla:2010cs,Radford:2009qi,Eichten:1974af,Eichten:1978tg} and the calculated results can be found in Tables \ref{tab:e1},\ref{tab:e11} and \ref{tab:m1}.
\begin{eqnarray}
 \Gamma(n^{2S+1}L_{iJ_i} \to n^{2S+1}L_{fJ_f} + \gamma) &=& \frac{4 \alpha_e \langle e_Q\rangle ^2\omega^3}{3} (2 J_f + 1) S_{if}^{E1} |M_{if}^{E1}|^2
 \end{eqnarray}
\begin{eqnarray}
\Gamma(n^3S_1 \to {n'}^{1}S_0+ \gamma) = \frac{\alpha_e \mu^2 \omega^3}{3} (2 J_f + 1) S_{if}^{M1} |M_{if}^{M1}|^2
\end{eqnarray}
Here, $\langle e_Q \rangle$, $\mu$ and $\omega$ are mean charge content, magnetic dipole moment and photon energy of the system respectively, and are written as,\\
%where, mean charge content $\langle e_Q \rangle$ of the system, magnetic dipole moment $\mu$ and photon energy $\omega$ are given by
\begin{equation}
\langle e_Q \rangle = \left |\frac{m_{\bar{q}} e_Q - e_{\bar{q}} m_Q}{m_Q + m_{\bar{q}}}\right |%|\frac{m_{\bar{Q}} e_Q - m_Q e_{\bar{Q}}}{m_Q + m_{\bar{Q}}|,
\end{equation}
\begin{equation}
\mu=\frac{m_{\bar{q}} e_Q - e_{\bar{q}} m_Q}{m_Q m_{\bar{q}}}
%\mu = \frac{e_Q}{m_\bar{q}} - \frac{e_{\bar{Q}}}{{m_Q}}
\end{equation}
\begin{equation}
\omega = \frac{M_i^2 - M_f^2}{2 M_i}
\end{equation}

\begin{equation}
S_{if}^{E1} = {\rm max}(L_i, L_f)
\left\{ \begin{array}{ccc} J_i & 1 & J_f \\ L_f & S & L_i \end{array} \right\}^2\\
\end{equation}
and
\begin{equation}
S_{if}^{M1} = 6 (2 S_i + 1) (2 S_f + 1)
\left\{ \begin{array}{ccc} J_i & 1 & J_f \\ S_f & \ell & S_i \end{array} \right\}^2 \left\{ \begin{array}{ccc} 1 & \frac{1}{2} & \frac{1}{2} \\ \frac{1}{2} & S_f & S_i \end{array} \right\}^2.
\end{equation}
The matrix element $|M_{if}|$ for $E1$ and $M1$ transitions can be written as
\begin{equation}
\left |M_{if}^{E1}\right | = \frac{3}{\omega} \left\langle f \left | \frac{\omega r}{2} j_0 \left(\frac{\omega r}{2}\right) - j_1 \left(\frac{\omega r}{2}\right) \right | i \right\rangle
\end{equation}
and
\begin{equation}
\left |M_{if}^{M1}\right | = \left\langle f\left | j_0 \left(\frac{\omega r}{2}\right) \right | i \right\rangle
\end{equation}

\begin{table*}[!h]
\begin{center}
\textit{\caption{\label{tab:e1} {E1 transition of $B$ meson.}}}
\vspace{0.1cm}
\resizebox{\textwidth}{!}{
\begin{tabular}{c c c c c c c c c c c c}
\hline
 State && Recent study &&  &&& Other study ($\Gamma$ in $keV$)&  \\
\hline
 Intial & Final & $ E_{\gamma}(MeV) $ & $\Gamma(keV)$ &  & \cite{Kher:2017mky}&\cite{Lu:2016bbk} &\cite{Godfrey:2016nwn}&\cite{PhysRevD.84.074030}& \\
 \hline
 $B(1^3P_{2})$&$B(1^3S_{1})$&437&315.054&&215.39  &177.7&444&131.36&\\
  $B(1^3P_{1})$&$B(1^3S_{1})$&429&316.32&  &227.09&108.5&300&122.87\\
  $B(1^1P_{1})$&$B(1^1S_{0})$&460&390.51&  &74.66&60.4&132&179.35\\
   $B(1^1P_{1})$&$B(1^3S_{1})$&426&310&  &51.90&53.1&97.5&8.98\\
  $B(1^1P_{1})$&$B(1^1S_{0})$&458&383&  &252.16&130.2&373&6.09\\
  $B(1^3P_{0})$&$B(1^3S_{1})$&423&302&  &201.34&116.9&325&100.54\\
  \hline
  $B(2^3S_{1})$&$B(1^3P_{2})$&201&37.89&  &34.75&51.6&30.8&10.33\\
  $B(2^3S_{1})$&$B(1^3P_{1})$&177&15.49&  &16.77&25.9&13.7&2.07\\
  $B(2^3S_{1})$&$B(1^1P_{1})$&180&16.26&  &5.82&11.67&5.32&13.79\\
  $B(2^3S_{1})$&$B(1^3P_{0})$&184&5.77&  &7.94&21.4&8.25&0.94\\
  $B(2^1S_{0})$&$B(1^3P_{1})$&168&39.93&  &9.43&25.2&9.01&\\
  $B(2^1S_{0})$&$B(1^1P_{1})$&171&42.03&  &49.39&49.6&31.1&\\
    \hline
  $B(1^3D_{3})$&$B(1^3P_{2})$&369&641.88&  &411.7&127&464&\\
  $B(1^1D_{2})$&$B(1^3P_{2})$&333&118&  &26.74&57.2&80.5&\\
  $B(1^1D_{2})$&$B(1^3P_{2})$&296&83.03&  &59.25&14.5&42.2&\\
  $B(1^3D_{1})$&$B(1^3P_{2})$&293&8.96&  &5.94&9.3&13&\\
  $B(1^3D_{1})$&$B(1^3P_{1})$&269&104.2&  &73.97&106.2&144&\\
  $B(1^3D_{1})$&$B(1^3P_{1})$&272&107.6&  &296.6&49&52.2&\\
  $B(1^3D_{1})$&$B(1^3P_{0})$&276&149.6&&131.38  &283.5&297&&\\
  $B(1^3D_{2})$&$B(1^3P_{1})$&309&379&  &153.8&356.3&433&\\
  $B(1^3D_{2})$&$B(1^1P_{1})$&312&389&  &17.6&8.6&29&\\
  $B(1^1D_{2})$&$B(1^1P_{1})$&272&258&  &10.4&0.1&2.67&\\
  $B(1^1D_{2})$&$B(1^1P_{1})$&275&266.6&  &133.2&143.1&397&\\
  \hline
  $B(2^3P_{2})$&$B(2^3S_{1})$&318&&  &319.93&&258&\\
  $B(2^3P_{1})$&$B(2^3S_{1})$&328&&  &340.32&&243&\\
  $B(2^3P_{1})$&$B(2^1S_{0})$&337&&  &30.47&&111&\\
  $B(2^1P_{1})$&$B(2^3S_{1})$&327&&  &83.99&&70.5&\\
  $B(2^1P_{1})$&$B(2^1S_{0})$&335&&  &35.3&&208&\\
  $B(2^3P_{0})$&$B(2^3S_{1})$&325&&  &31.14&&66.7&\\
    \hline
  $B(2^3P_{2})$&$B(1^3D_{3})$&149&0.6&  &32.26&&16.5&&\\
  $B(2^3P_{2})$&$B(1^3D_{2})$&186&5.21&  &7.22&&&\\
  $B(2^3P_{2})$&$B(1^1D_{2})$&223&9.07&  &3.05&&14.8&\\
  $B(2^3P_{2})$&$B(1^3D_{1})$&226&2.03&  &0.88&&&\\
  $B(2^3P_{1})$&$B(1^3D_{1})$&237&11.67&  &6.73&&3.68&\\
  $B(2^1P_{1})$&$B(1^3D_{1})$&235&11.39&  &20.31&&&\\
  $B(2^3P_{0})$&$B(1^3D_{1})$&233&14.82&  &86.17&&16.1&\\
   \hline
\end{tabular}
}
\end{center}
\end{table*}

\begin{table*}[!h]
\begin{center}
\textit{\caption{\label{tab:e11} {E1 transition of $B_{s}$ meson.}}}
\vspace{0.1cm}
\resizebox{\textwidth}{!}{
\begin{tabular}{c c c c c c c c c c c c}
\hline
 State && Recent study &&  &&& Other study ($\Gamma$ in $keV$)&  \\
\hline
 Intial & Final & $ E_{\gamma}(MeV) $ & $\Gamma(keV)$ &  & \cite{Kher:2017mky}&\cite{Lu:2016bbk} &\cite{Godfrey:2016nwn}&\cite{PhysRevD.84.074030}& \\
 \hline
 $Bs(1^3P_{2})$&$Bs(1^3S_{1})$&407&81.73&  &150.59&159&106.0&131.36\\
  $Bs(1^3P_{1})$&$Bs(1^3S_{1})$&397&80.62&  &151.14&98.8&57.3&122.87\\
  $Bs(1^3P_{1})$&$Bs(1^1S_{0})$&439&108.71&  &35.59&56.6&47.8&179.35\\
   $Bs(1^1P_{1})$&$Bs(1^3S_{1})$&396&80.05&  &24.64&39.5&36.9&8.98\\
  $Bs(1^1P_{1})$&$Bs(1^1S_{0})$&438&108.03&  &184.92&97.7&70.6&06.09\\
  $Bs(1^3P_{0})$&$Bs(1^3S_{1})$&375&68.25&  &123.75&84.7&76.0&100.54\\
  \hline
  $Bs(2^3S_{1})$&$Bs(1^3P_{2})$&151&5.80&  &18.81&25.6&8.08&10.33\\
  $Bs(2^3S_{1})$&$Bs(1^3P_{1})$&152&3.61&  &1.069&1.6&3.2&2.07\\
  $Bs(2^3S_{1})$&$Bs(1^1P_{1})$&153&3.68&  &2.34&9.4&2.28&13.79\\
  $Bs(2^3S_{1})$&$Bs(1^3P_{0})$&175&1.81&  &5.64&17.2&2.52&0.94\\
  $Bs(2^1S_{0})$&$Bs(1^1P_{1})$&142&8.72&  &3.58&12.3&4.66&\\
  $Bs(2^1S_{0})$&$Bs(1^1P_{1})$&143&8.90&  &26.37&41.7&6.73&\\
    \hline
  $Bs(1^3D_{3})$&$Bs(1^3P_{2})$&307&128&  &249.92&113.2&109.6&\\
  $Bs(1^3D_{2})$&$Bs(1^3P_{2})$&286&25.94&  &16.64&31.5&18.8&\\
  $Bs(1^1D_{2})$&$Bs(1^3P_{2})$&261&19.56&  &38.93&10.5&10.2&\\
  $Bs(1^3D_{1})$&$Bs(1^3P_{2})$&261&2.17&  &4.14&5.1&3.07&\\
  $Bs(1^3D_{1})$&$Bs(1^3P_{1})$&263&33.32&  &57.35&56.8&28.5&\\
  $Bs(1^3D_{1})$&$Bs(1^1P_{1})$&264&33.69&  &21.46&35.3&19.6&\\
  $Bs(1^3D_{1})$&$Bs(1^3P_{0})$&285&56.50&  &109.42&204.4&74.7&\\
  $Bs(1^3D_{2})$&$Bs(1^3P_{1})$&288&105.8&  &151.20&195.4&112.0&\\
  $Bs(1^3D_{2})$&$Bs(1^1P_{1})$&289&106.9&  &17.01&5.9&1.07&\\
  $Bs(1^1D_{2})$&$Bs(1^3P_{1})$&263&79.98&  &1.18&0.04&0.368&\\
  $Bs(1^1D_{2})$&$Bs(1^1P_{1})$&264&80.85&  &131.42&138.8&95.9&\\
  \hline
  $Bs(2^3P_{2})$&$Bs(2^3S_{1})$&305&86.74&  &211.55&&65.4&\\
  $Bs(2^3P_{1})$&$Bs(2^3S_{1})$&305&86.74&  &212.6&&52.3&\\
  $Bs(2^3P_{1})$&$Bs(2^1S_{0})$&315&95.95&  &12.51&&25.2&\\
  $Bs(2^1P_{1})$&$Bs(2^3S_{1})$&305&86.74&  &42.78&&18.6&\\
  $Bs(2^1P_{1})$&$Bs(2^1S_{0})$&315&95.95&  &237.2&&50.7&\\
  $Bs(2^3P_{0})$&$Bs(2^3S_{1})$&291&111&  &186.64&&66.8&\\
    \hline
  $Bs(2^3P_{2})$&$Bs(1^3D_{3})$&195&11.0&  &21.90&&5.61&&\\
  $Bs(2^3P_{2})$&$Bs(1^1D_{2})$&169&1.277&  &4.72&&&\\
  $Bs(2^3P_{2})$&$Bs(1^1D_{2})$&195&1.96&  &1.49&&&\\
  $Bs(2^3P_{2})$&$Bs(1^3D_{1})$&148&0.056&  &0.51&&&\\
  $Bs(2^3P_{1})$&$Bs(1^3D_{1})$&195&3.27&  &2.90&&0.944&\\
  $Bs(2^1P_{1})$&$Bs(1^3D_{1})$&195&3.27&  &11.51&&&\\
  $Bs(2^3P_{0})$&$Bs(1^3D_{1})$&182&10.56&  &42.98&&3.87&\\
   \hline
\end{tabular}
}
\end{center}
\end{table*}

\begin{table*}[!h]
\begin{center}
\textit{\caption{\label{tab:m1}{M1 transition of $B$ and $B_{s}$ mesons.}}}
\vspace{0.1cm}
\resizebox{\textwidth}{!}{
\begin{tabular}{c c c c c c c c c c c c}
\hline
 State& Recent study &&  &&&Other study ($\Gamma$ in $keV$) \\
\hline
  Intial$\rightarrow$ Final & $ E_{\gamma}(MeV) $ & $\Gamma(keV)$ & \cite{tanabashi2018particle} & \cite{Kher:2017mky}&\cite{Lu:2016bbk} &\cite{Godfrey:2016nwn}& \cite{PhysRevD.84.074030}& \cite{ebert2002radiative}&\cite{bhatnagar2020radiative}&\cite{choi2007decay}\\
 \hline
 $B(1^3S_{1})$ $\rightarrow$ $B(1^1S_{0})$&41&0.135&0.13$ \pm 0.01 $&0.069&0.1&1.23&1.258& 0.19&0.1472&0.13$ \pm 0.01$\\
 $B(2^3S_{1})$ $\rightarrow$ $B(2^1S_{0})$&8.99&0.013&&0.014&0.05&&0.018&&0.1458\\
 $B(3^3S_{1})$ $\rightarrow$ $B(3^1S_{0})$&3.99&0.11&&0.005&&&0.003\\
  $B(2^3S_{1})$ $\rightarrow$ $B(1^1S_{0})$&624&65.078&&53.79&8.0&67.4&\\
   $B(2^1S_{0})$ $\rightarrow$ $B(1^3S_{1})$&616&187.58&&124.1&0.9&108&\\
 \hline
  $B(1^1P_{1})$ $\rightarrow$ $B(1^3P_{0})$&3.99&0.0001&&0.00003&0.03&&\\
  $B(1^1P_{1})$ $\rightarrow$ $B(1^3P_{0})$&6.99&0.000633&&0.017&0.01&&\\
   $B(1^{3}P_{2})$ $\rightarrow$ $B(1P_{1})$ &2&&&0.002&0.0014&&\\
   $B(1^1P_{1})$ $\rightarrow$ $B(1^3P_{2})$&2&&&0.015&0.017&&\\
  \hline
$B(1^{3}D_{3})$ $\rightarrow$ $B(1^1D_{2})$&2.99&0.000249&&1.29&0.7&&\\
$B(1^{3}D_{3})$ $\rightarrow$ $B(1^{3}D_{1})$&38.87&&&1.479&&&\\
$B(1^3D_{2})$ $\rightarrow$ $B(1^{3}D_{1)}$&41.8&0.402&&0.542&0.0023&&\\
$B(1^3D_{2})$ $\rightarrow$ $B(1^1D_{2})$&38.8&0.538&&0.314&1.9&&\\
  \hline
   $Bs(1^3S_{1})$ $\rightarrow$ $Bs(1^1S_{0})$&55&0.102&0.064$ \pm 0.016 $&0.095&0.1&0.313&0.286&0.054\\
 $Bs(2^3S_{1})$ $\rightarrow$ $Bs(2^1S_{0})$&10&0.00079&&0.018&0.02&&0.008&&0.0531\\
 $Bs(3^3S_{1})$ $\rightarrow$ $Bs(3^1S_{0})$&4.99&0.0000744&&0.007&&&0.001\\
  $Bs(2^3S_{1})$ $\rightarrow$ $Bs(1^1S_{0})$&580&16.92&&35.06&4&14.2&\\
   $Bs(2^1S_{0})$ $\rightarrow$ $Bs(1^3S_{1})$&530&38.20&&85.03&0.1&3.23&\\
 \hline
  $Bs(1^1P_{1})$ $\rightarrow$ $Bs(1^3P_{0})$&2.19&0.063&&0.013&0.02&&\\
  $Bs(1^1P_{1})$ $\rightarrow$ $Bs(1^3P_{0})$&2.29&0.0072&&0.029&0.05&&\\
   $Bs(1^{3}P_{2})$ $\rightarrow$ $Bs(1^1P_{1})$ &2.99&0.00004&&0.005&0.04&&\\
   $Bs(1^1P_{1})$ $\rightarrow$ $Bs(1^3P_{2})$&1.99&0.0000121&&0.00011&0.000052&&\\
  \hline
$Bs(1^{3}D_{3})$ $\rightarrow$ $Bs(1^1D_{2})$&0.9&&&0.445&0.1&&\\
$Bs(1^{3}D_{3})$ $\rightarrow$ $Bs(1^{3}D_{1})$&0.4&0.206&&0.462&&&\\
$Bs(1^1D_{2})$ $\rightarrow$ $Bs(1^{3}D_{1})$&26.94&0.0348&&0.168&0.0013&&\\
$Bs(1^3D_{2})$ $\rightarrow$ $Bs(1D_{2})$&26.94&0.0581&&0.125&0.4&&\\
  \hline
 \end{tabular}
 }
\end{center}
\end{table*}

\section{Results, Discussion and Conclusion}
\label{sec:Result}
In present article we have calculated the $B$ and $B_s$ meson spectrum in semi-relativistic approach. We have added relativistic correction to the kinetic energy and potential energy terms in the potential, for the present work we have used the screening potential. Numerical values of various orders of relativistic corrections for $B$ and $B_s$ meson can be found in Table \ref{tab:p orders}. The Gaussian like wave-function in position and momentum space has been used, and the potential has been solved using the variational approach. The spin average masses and the masses for $s,p,d \& f$ states of $B$ and $B_s$ meson can be found in Tables \ref{tab:SA},\ref{tab:Bmass},\ref{tab:Bmass1},\ref{tab:Bsmass},\ref{tab:Bsmass1}, respectively. On analyzing the masses, it is evident that for states $n \geq 3$ the spin-average masses for the present study are suppressed in compression to \cite{Kher:2017mky}(semi-relativistic approach, but using Cornel potential). The masses calculated in the present article are consistent with the masses calculated by Relativistic quark model\cite{Ebert:2009ua}, Godfrey Isgur model\cite{Godfrey:2016nwn}, Godfrey Isgur model along with screened potential model\cite{Ebert_2010}, and R.Q.M based on a heavy-quark expansion of the instantaneous Bethe-Salpeter equation\cite{Liu:2016efm}.\\

It is well proven fact that ${{\mathit B}_{{J}}^{*}{(5732)}}$, first observed at OPAL detector, later separated into two states ${{\boldsymbol B}_{{1}}{(5721)}^0}$ and ${{\mathit B}_{{2}}^{*}{(5747)}^0}$.
For $B$ meson we juxtapose the calculated masses from present approach with experimentally determined masses and masses from other theoretical approaches, as can be seen in Tables \ref{tab:Bmass} \& \ref{tab:Bmass1}. We observe that the two states $B^{0}$ and $B^{*}$ are reproduced well with only a difference of $3$ and $0$ Mev, respectively in comparison with PDG masses\cite{tanabashi2018particle}. As per theoretical studies by \cite{Chen:2016spr,Yu:2019iwm,li:2021hss,Kher:2017mky} ${{\boldsymbol B}_{{1}}{(5721)}^0}$ has been regarded as ${{\boldsymbol B}(1^1 P_1)}$ or ${{\boldsymbol B}(1^3 P_1)}$ state or as an admixture of these two states. But, depending upon our calculated masses and constructed $(J,M^2)$ Regge trajectory in Fig.\ref{fig:un} we associate it as $1^1 P_1$ state, because the calculated mass from the present study show linearity and parallelism and fit well on the parent $(J,M^2)$ trajectory. For ${{\mathit B}_{{2}}^{*}{(5747)}^0}$ state, the difference between mass calculated by us and that from PDG\cite{tanabashi2018particle} is only $0.7 \%$ and the constructed $(J,M^2)$ Regge trajectory in Fig.\ref{fig:natural} helps in associating ${{\mathit B}_{{2}}^{*}{(5747)}^0}$ as $1^3 P_2$ state. It is well understood that ${{\boldsymbol B}_{{J}}{(5960)}}$ and ${{\boldsymbol B}_{{J}}{(5970)}}$ are same states,and their masses are nearby the estimated masses of $2^1S_0$ and $2^3S_1$ states. Because, ${{\boldsymbol B}_{{J}}{(5970)}}$ decays to $B\pi$, we eliminate its association to $2^1S_0$  state. Our calculated mass for ${{\boldsymbol B}_{{J}}{(5970)}}$ differs from PDG value by $2$ Mev only. Thus, helping us relate ${{\boldsymbol B}_{{J}}{(5970)}}$ as $2^3S_1$ state. Also, our present study help us link ${{\mathit B}_{{J}}{(5840)}^0}$ with $2^1S_0$ state. The $(n_r, M^2)$ Regge trajectory in Fig.\ref{fig:ps} show linearity and parallelism with both ${{\boldsymbol B}_{{J}}{(5970)}}$ and ${{\mathit B}_{{J}}{(5840)}^0}$ lying on pseudoscalar and vector trajectories, resulting in validating our association of ${{\boldsymbol B}_{{J}}{(5970)}}$ and ${{\mathit B}_{{J}}{(5840)}^0}$ as $2^3S_1$ and $2^1S_0$ states.\\

In the $B_s$ meson family two $1S$ states $B_{S}$ and $B^{*}_{S}$ are reproduced well. the difference between our calculated masses and that from PDG for these two states differ only by $7$ and $0$ Mev respectively. ${{\mathit B}_{{sJ}}^{*}{(5850)}}$ state was first observed at OPAL detector, was later severed into two different states ${{\mathit B}_{{s1}}{(5830)}^{0}}$ and ${{\mathit B}_{{s2}}^{*}{(5840)}^{0}}$. Based on our study of mass spectra and constructed Regge trajectories in Fig.\ref{fig:natural1} and \ref{fig:un1} we associate ${{\mathit B}_{{s1}}{(5830)}^{0}}$ as $1 ^1P_1$ state and ${{\mathit B}_{{s2}}^{*}{(5840)}^{0}}$ as $1^3P_2$ state. Calculated masses from current study fall within the error margin of the masses determined by Expt.(PDG)\cite{tanabashi2018particle}, and also match well with other theoretical studies. Two new states ${\mathit B}_{{SJ}}{(6064)}$ and ${\mathit B}_{{SJ}}{(6114)}$ were recently observed at LHCb\cite{LHCb:2020pet} in the $B^+ K^-$ mass spectrum at a mass approximately $300$ Mev above $B^+ K^-$ threshold. As per mass predictions from various quark models these two new states can be good candidates for $1D$ wave states. The calculated mass for $1^1D_2$ state for $B_s$ meson from the current study varies only by $0.67 \%$ with respect to the experiment mass from LHCb for ${\mathit B}_{{SJ}}{(6064)}$ state. The calculated mass fits very well on the parent $(J,M^2)$ Regge trajectory and follows linearity and parallelism. Hence, we associate ${\mathit B}_{{SJ}}{(6064)}$ with $1^1D_2$ $B_s$ meson state.\\
According to our analysis, we interpret ${\mathit B}_{{SJ}}{(6114)}$ as an admixture of $2^3S_1$ ($5995$ Mev)and $1^3D_1 $ ($6112$ Mev) considering mixing angle $\theta = 85 ^\circ$ we calculate mass of ${\mathit B}_{{SJ}}{(6064)}$ as $6111.1$ Mev which lies in experimentally determined error bar $6114 \pm 5 $ Mev \cite{LHCb:2020pet}. More experimental and theoretical analysis is required to throw more light on these two newly observed $B_s$ meson states. We also predict $s,p,d$ \& $f$ states for both $B$ and $B_s$ mesons and compare with available theoretical data.\\
The decay constant for $s$-wave states for both $B$ and $B_s$ meson have been calculated using the Van-Royen-Weisskoph formula, the first order QCD correction factor has also been incorporated. The calculated decay constant can be found in Table \ref{tab:decayconstant}. The calculated decay constant considering QCD correction are bit suppressed when compared to decay constant values calculated experimentally.\\
The calculated values of branching fraction of the $B$ meson is tabulated in Table \ref{tab:tableBF}, and compared with theoretically and experimentally determined values. For $Br_{\mu}$ and $Br_{e}$ our predictions are in accordance with experimentally determined branching fraction values, but for $Br_{\tau}$ the calculated value is suppressed.\\
Theoretically many methods are available to calculate radiative leptonic decay width and branching ratio, like in Ref.\cite{Kher:2017mky} considering non-relativistic quark model, ref.\cite{Korchemsky:1999qb} considering perturbative QCD approach, factorization approach in \cite{Yang:2011ie,Yang:2014rna,Yang:2016wtm}. In all these approaches the value of branching ratio is in the range of $10^{-6}$ for $B$ meson. In present study the determined value of branching ratio is of the order $10^{-7}$ for both the $B$ and $B_s$ meson, the calculated values of branching ratio with corresponding radiative leptonic decay width can be found in Table \ref{tab:tablerld}.\\
In the Tables \ref{tab:rare1} \& \ref{rare2} is tabulated predicted branching ratio with corresponding rare leptonic $(B^{0}_{q}\rightarrow l^{+} l^{-})$ decay width for both the $B$ and $B_s$ meson. The predicted branching ratios for $B^{0}\rightarrow\mu^{+}\mu^{-}$ and $B^{0}_{s}\rightarrow\mu^{+}\mu^{-}$ are in good agreement with CMS and LHCb\cite{CMS:2013dcn,LHCb:2013vgu,CMS:2014xfa}. The predicted branching ratio for $ B^{0}\rightarrow\tau^{+}\tau^{-}$ and $ B^{0}_{s}\rightarrow\tau^{+}\tau^{-}$ agree well with experimental an other theoretical values. But, the branching ratio for $ B^{0}\rightarrow e^{+}e^{-}$ and $ B^{0}_{s}\rightarrow e^{+}e^{-}$ also agree well with theoretical predications.\\
We have also calculated various mixing parameters for both $B$ and $B_s$ meson. Our predicated mass difference $\bigtriangleup m_{B_d}$ ($0.253$ ps^{-1}) and $\bigtriangleup m_{B_d}$ ($25.37$ ps^{-1}) is not away from the experimentally determined mass differences. The calculated value of the mixing parameter $\chi_{q}$  for $B$ and $B_s$ meson is $0.415$ and $12.68$ respectively, the experimental value for mixing parameter are $0.770\pm0.004$ and $26.79\pm0.08$ respectively. Other mixing parameters $x_{q}$ and $y_{q}$ can also be found in Table\ref{tab:mix}. The values of the mixing parameters are suppressed because in the present article we have only considered short distance contributions.\\
Employing calculated masses and normalised reduced wave-functions the electromagnetic transition widths for $B$ and $B_s$ meson have also been calculated and tabulated in Tables\ref{tab:e1}, \ref{tab:e11} \& \ref{tab:m1}. The calculated electromagnetic transition width are compared with experimental and other theoretical widths. \\
Finally, we conclude that the masses of $B$ and $B_s$ mesons determined in present article using a semi-relativistic approach and considering screening potential are in fine tune with experimental masses wherever applicable and are also comparable with masses from other theoretical approaches. The screening potential suppresses the masses for states with $n \geq 3$. The mass spectroscopy and $(n_r,M^2)$, $(J,M^2)$ Regge trajectories helps us associate ${{\boldsymbol B}_{{1}}{(5721)}^0}$, ${{\mathit B}_{{2}}^{*}{(5747)}^0}$, ${{\boldsymbol B}_{{J}}{(5970)}}$, and ${{\mathit B}_{{J}}{(5840)}^0}$  in the $B$ meson family as $1 P_1$, $1^3 P_2$, $2^3S_1$ and $2^1S_0$ states. In the $B_s$ meson family we associate ${{\mathit B}_{{s1}}{(5830)}^{0}}$, ${{\mathit B}_{{s2}}^{*}{(5840)}^{0}}$, ${\mathit B}_{{SJ}}{(6064)}$ as $1 ^1P_1$, $1^3P_2$, and $1^1D_2$ states, and ${\mathit B}_{{SJ}}{(6114)}$ as an admixture of $2^3S_1$ and $1^3D_1 $ states.

\label{sec:Conclusion}

\bibliographystyle{spphys}
\bibliography{epjc}

\end{document}